\documentclass[
  aps,
  prd,
  reprint,
  groupedaddress,
  nofootinbib,
  nobibnotes,
  amsmath,
  amssymb,
  ]{revtex4-1}

\usepackage{epsfig}
\usepackage{graphicx}
\usepackage{url}
\def\w{\omega}
\def\o{\phi}
\def\<{\langle}
\def\>{\rangle}
\def\curl{\mathcal}
\def\ell{l}
\def\un{\hat{n}}

\begin{document}

\title{Combining power spectrum and bispectrum measurements to detect oscillatory features}

\author{J.R.~Fergusson$^{1}$}
\email{j.fergusson@damtp.cam.ac.uk}
\author{H.F.~Gruetjen$^{1}$}
\email{hfg22@cam.ac.uk}
\author{E.P.S.~Shellard$^{1}$}
\author{M.~Liguori$^{2}$}

\affiliation{$^{1}$Centre for Theoretical Cosmology, DAMTP, University of Cambridge, Cambridge CB3 0WA, United Kingdom}
\affiliation{$^{2}$Dipartimento di Fisica e Astronomia G. Galilei,Universit\`a degli Studi di Padova, via Marzolo 8, I-35131 Padova, Italy}

\date{\today}

\begin{abstract}
The simplest inflationary models present us with few observable parameters to discriminate between them. A detection of features in the spectra of primordial density perturbations could provide valuable insights and lead to stringent tests of models of the early Universe. So far, searches for oscillatory features have not produced statistically significant results. In this work we consider a combined search for features in the power spectrum and bispectrum. We show that possible dependencies between the estimates of feature model amplitudes based on the two- and three-point correlators are largely statistically independent under the assumption of the null hypothesis of a nearly Gaussian featureless cosmic microwave background. Building on this conclusion we propose an optimal amplitude estimator for a combined search and study the look-elsewhere effect in feature model surveys. In particular we construct analytic models for the distribution of amplitude estimates that allow for a reliable assessment of the significance of potential findings. We also propose a well-behaved integrated statistic that is designed to detect evidence for models exhibiting features at multiple frequencies.
\end{abstract}

\maketitle
\tableofcontents

\section{Introduction}
The recent Planck results have further cemented the place of the inflationary paradigm as the best explanation of how our Universe began. The key observational predictions of this model are flat, isotropic and homogeneous universe with an approximately scale-invariant primordial power spectrum. Traditionally there are only two observable parameters which we can use to discriminate between differing inflationary models, the spectral tilt, $n_s$, and the tensor to scalar ratio, $r$.  Despite the observational evidence that $n_s \approx 0.96$ (see Ref.~\cite{Ade:2013CosmoParam}), this constraint has not proved to be a significant barrier to model building with a plethora of viable candidates. The possible detection of $r$ could have a more decisive impact but this has yet to be verified \cite{Ade:2014xna, Adam:2014bub}. For this reason attention has also been focussed on other observables which may be able to differentiate between models.
One of the most promising is non-Gaussianity. The simplest slow-roll single-field inflationary models predict the primordial density fluctuations to be Gaussian to a high degree but this so-called standard model is arguably poorly motivated in fundamental theory.  On the other hand, more realistic inflationary models can produce characteristic non-Gaussian signals whose form is closely related to the specific dynamics underlying the theory.

Non-Gaussianity is commonly constrained by measuring the bispectrum (the Fourier transform of the three-point correlator) of the cosmic microwave background (CMB). This was the focus of the recent Planck cosmology paper on non-Gaussianity \cite{Ade:2013NonGaussianity}, which found no evidence for a significant bispectrum for a wide range of scale-invariant non-Gaussian models. The only 'hints' of deviations from Gaussianity were those with a non-Bunch-Davies (NBD) or excited initial vacuum state and those with oscillatory-type modulations, both observed at a significance level of about  2$\sigma$ (see also earlier feature model searches in the WMAP bispectrum \cite{Fergusson:Bispectrum2010}). While we will have to wait for improved  data sets to improve constraints on NBD-type models, oscillatory-type models predict perturbations to the power spectrum which could also be observed. This was first discussed for models that arise from features in the inflationary potential in \cite{Chen:LargeNonGaussanities} and for models where there is a resonance with oscillatory features in the inflationary potential in \cite{Chen:Generation}. Subsequently, detailed studies of the effects of features in the inflationary potential and the speed of sound have been undertaken \cite{Dvorkin:GSR, Adshead:NonGaussianity, Miranda:WarpFeatures, Achucarro:SoundFeatures,Bartolo:EFTfeatures}. It has been shown that for a wide range of parameter space the power spectrum  and the bispectrum oscillate with the same underlying frequency. This special frequency relationship between polyspectra has been found to be a robust property of oscillatory-type models; see the review article \cite{Chen:PrimNonGaussianities} and subsequent references \cite{Flauger:Resonant, Chen:FoldedResonant, Arroja:LargeStrong, Martin:ScalarBispectrum, Adshead:NonGaussianity, Achucarro:CorrelatingFeatures, Bartolo:EFTfeatures, Gong:CorrelatingCorrelation}. We can approximate it with the following relation between the power spectrum and the bispectrum:
\begin{eqnarray}
\label{eq:exemplarPS}
P_{\mathcal{R}}(k) &=& P_{\mathcal{R},0}(k)\left(1+A_P\sin(\omega (2k) + \phi_P)\right),\\
\label{eq:exemplarBS}
B(k_1,k_2,k_3) &=&\frac{A_B \Delta^4_{\mathcal{R}}(k_{\star})}{(k_1 k_2 k_3)^2}  \sin(\omega K + \phi_B)\,,
\end{eqnarray}
where $K=k_1+k_2+k_3$, $P_{R,0}(k)$ is the power spectrum in the absence of any feature, $\Delta^2_{R}(k)=k^3/(2\pi^2)P_{R,0}(k)$ is the dimensionless power spectrum and $k_{\star}$ is a fiducial momentum scale. Here, we define five model-dependent parameters, the common frequency $\omega$, the relative amplitudes $A_P,\;A_B$ and the relative phases $\o_P,\o_B$. This is our exemplar model for which we will mainly consider a generic frequency signal search with unknown relations between the feature polyspectra amplitudes and phases, though we will also consider the implications of fixing both the amplitude ratio $A_B/A_P$ and the relative phase $\o_B-\o_P$. Many searches for similar oscillatory features in the CMB power spectrum based on WMAP and Planck data have been performed elsewhere (see for example \cite{Pahud:OscInflaton,Meerburg:2011WMAP7Constraints,Peiris:2013ConstMonoInf,Flauger:2009MonoInf,Meerburg:2013SearchOscP1,Meerburg:2013SearchOscP2,Achucarro:SearchCsFeatures, Meerburg:2014SearchOsc}). Our primary purpose is to discuss the implications of positive results at the same frequency for both the power spectrum and bispectrum as evidence for this type of model.  
We note that specific oscillatory models will typically have additional parameters when compared with (\ref{eq:exemplarPS}) and (\ref{eq:exemplarBS}), such as those which define the feature signal envelope.  So apart from the additional free parameters, they will generically have a narrower effective domain over which they can be detected, with less signal to noise (S/N) at a given amplitude. For this reason, our discussion of statistical issues, like the `look elsewhere' effect, using the simple feature model defined above with 3-5 parameters should be considered conservative; more complicated models may have to cross a higher threshold in terms of raw statistical significance. 

Despite the apparent simplicity of the exemplar model, we emphasise that it is well motivated physically. In the context of single-field slow-roll models, it is possible to quantitatively predict the periodic excitations in all polyspectra caused by features in the slow-roll parameters (see, for example, Ref.~\cite{Adshead:NonGaussianity}). These models typically have a damping envelope and, in the sharp feature limit, the damping weakens so the solution approximates%
\footnote{For simplicity, we ignore possible scalings of the oscillation amplitudes with wave number in this work. In the case of the power spectrum the leading order behaviour is usually a constant cosine oscillation like in our exemplar model, while the sine mode comes with a factor of $1/(k\omega)$. Typically feature models produce a bispectrum with a cosine oscillation of amplitude $~(K\omega)^2$, while the sine oscillation only scales as $K\omega$. Despite the scalings, these oscillations are clearly highly correlated with the sine and cosine templates of our simple model.} %
(\ref{eq:exemplarPS}) and (\ref{eq:exemplarBS}). The relative amplitude $A_B/A_P$ and other parameters are feature dependent. Hence, the detectability of the bispectrum relative to the power spectrum depends on the particular feature model under consideration.

Feature signals arising from multifield models are even more model dependent, in particular in terms of the relative power spectrum and bispectrum amplitudes. However, there are interesting special cases in certain limits that have been studied and make definite predictions.   In Ref.~\cite{Achucarro:CorrelatingFeatures}, it was shown that the effect of additional heavy fields could be integrated out in certain limits, yielding a reduced sound speed. A nontrivial trajectory could induce features in the power spectrum, with a corresponding signal in the bispectrum calculated from the power spectrum and its two derivatives. In Ref.~\cite{Chen:PrimFeatures, Chen:StandardClock}, the idea of distinguishing inflation from other scenarios focussed on the possibility of exciting several heavy fields with a sharp feature;  these fields then oscillate around their minima creating several corresponding feature signals in both power spectrum and bispectrum. Like models in which the inflationary potential contains multiple features, this is a model in which multiple feature peaks could appear in the power spectrum and bispectrum, motivating our final sections which will discuss this case.   

Given that estimators for feature models based on the power spectrum and bispectrum are based on the same set of multipoles $a_{lm}$, an important aspect of a combined search is to exclude the possibility of dependencies between the constraints. Naively, if we consider a single multipole $a$, then the random variable $P:=a^2$ and $B:=a^3$ are clearly uncorrelated as $a$ is Gaussian with zero mean. However, $P$ and $B$ are obviously highly dependent and contain no complementary information (apart from the sign). Similarly, while we know that the estimators for connected parts of different correlators should be close to uncorrelated (as we know the $a_{lm}$ are approximately Gaussian) we also know that they are not entirely independent. A main focus of this paper will be to determine the degree to which estimators for oscillatory models based on different correlators are independent and can be combined to enhance our ability to detect feature models.

This paper is organised as follows. In Sec.~\ref{sec:Methods} we describe the simulation setup and pipelines used to study possible dependencies between estimates of feature model amplitudes in the power spectrum and bispectrum. Using standard measures of statistical dependence we go on to show that the measurements can be considered independent to a very good approximation in Sec.~\ref{sec:Results}. We also provide analytic arguments why this is expected in the large sample (large $l_{\text{max}}$) limit. Building on this result we study the implications for a combined search for feature models in Sec.~\ref{sec:CombSearch}. We develop analytic models for the distributions of amplitudes under the null hypothesis that allow us to judge the LE adjusted significances of potential findings. Section \ref{sec:RefinedSearch} extends these results to a survey allowing for feature models with multiple frequencies. In particular we suggest a simple integrated statistic that we found to provide a good assessment of whether a survey provides evidence for such multifrequency models. We also briefly discuss the possible inclusion of polarisation measurements and higher order correlators. We summarise and discuss our results in Sec.~\ref{sec:Results}.

\section{Methods}
\label{sec:Methods}
To answer the question whether or not there exist significant correlations between estimates of the amplitudes of feature models in the power spectrum and bispectrum it is necessary to obtain amplitudes for a large number of simulated CMB realisations. In the following sections we describe the two sets of simulations underlying our results and outline the pipelines used for the power spectrum and bispectrum analysis.

\subsection{CMB simulations}
To study whether there is any kind of correlation in the simplest scenario, we generate 10000 noiseless, unlensed full-sky CMB realisations of the Planck best-fit $\Lambda$CDM model. Each map is then processed by the two pipelines described in the following sections. We restrict the analysis in both cases to the multipole range $2\le l\le 2000$.

We also test whether incomplete sky coverage and instrumental noise can give rise to correlations. We generate 400 unlensed CMB realisations of the Planck best-fit $\Lambda$CDM model at a HEALPix resolution of $N_s=2048$. We multiply the multipoles by the beam window function for a 5 arcmin Gaussian beam and the pixel window function at resolution $N_s=2048$. Anisotropic coloured noise is obtained by first drawing a random number from a normal distribution for each pixel. The resulting noise realisations are then rescaled with an appropriate pixel variance map to introduce anisotropy. The resulting noise realisations have a flat unity power spectrum. To introduce correlations between pixels we transform to multipole space and rescale each multipole with a noise power spectrum $N_l$. Transforming back to pixel space we arrive at coloured anisotropic noise maps. By adding two different noise realisations to each CMB map we produce two maps for each of the 400 CMB realisations. We refer to these as half-noise (HN) maps. This allows us to employ cross-correlators between the two maps in the power spectrum analysis. These have the advantage that the noise does not lead to a bias and are preferable in practice over autocorrelators. The noise power spectrum and directional dependance is chosen such that the average of the two HN maps approximately mirrors the noise found in the Planck Spectral Matching Independent Component Analysis (SMICA) map \cite{Ade:2013CompSep}. This implies that each HN map is generated with twice the noise power spectrum extracted from the SMICA map.

The effect of incomplete sky coverage is incorporated by masking the maps with the union mask U73 \cite{Ade:2013CompSep, Ade:2013NonGaussianity} as well as two larger masks we constructed by extending the galactic cut of the U73 mask. The latter have sky coverage of 56\% and 38\% respectively and we refer to these modified union masks as the MU56 and MU38 masks.

\subsection{Power spectrum pipeline}
As has become standard we use a pseudo-$C_l$ (PCL) likelihood based on cross correlators to analyse the power spectrum \cite{Larson:WMAP7powerspectra, Ade:2013Likelihood, Ade:2013CosmoParam}. Below we outline how the procedure works in the more complicated case of the 400 pairs of HN maps. The analysis of the 10000 noiseless full-sky maps is simply obtained by setting the noise to zero, the mask to unity and replacing each HN map with the single realisation of the noiseless CMB.

To minimise leakage of power we apodise the masks by approximate convolution with a Gaussian beam of FWHM $.5^{\circ}$ using the procedure outlined in \cite{Gruetjen:PCL}. From the masked HN maps we extract the respective sets of multipole coefficients $a^1_{lm}$ and $a^2_{lm}$ and construct an unbiased power spectrum estimate given by
\begin{equation}
\hat{C}_{l_1}=M_{l_1l_2}^{-1}\tilde{C}_{l_2}\quad \tilde{C}_l=\frac{1}{2 l+1}\sum\limits_{m}a^1_{lm}a^2_{lm}
\end{equation}
where $M_{l_1l_2}$ is the standard PCL coupling matrix \cite{Hivon:MASTER}. We approximate the PCL log likelihood with the fiducial Gaussian approximation \cite{Efstathiou:MythsandTruths, HamimecheLewis:CMBlikelihoods, Ade:2013Likelihood}
\begin{equation}
-2\log{\cal{L}}=\left(\hat{C}_{l_1}-C_{l_1}\right)\Delta^{-1}_{l_1l_2}\left(\hat{C}_{l_2}-C_{l_2}\right)
\end{equation}
where $\Delta_{l_1l_2}=\langle \Delta\hat{C}_{l_1}\Delta\hat{C}_{l_2}\rangle$ is the covariance of the PCL estimates assuming the fiducial model is correct. We employ the analytic approximations from \cite{Efstathiou:MythsandTruths} to calculate the covariance matrices. These approximations assume an approximately constant power spectrum and thus do not properly account for leakage effects. This typically leads to an underestimate of the variance that we correct for using an improved analytic approximation \cite{Gruetjen:PCL}. Due to significant deviations from a Gaussian distribution at low $l$ the fiducial Gaussian approximation is not reliable in this region. We thus only consider the multipole range $50\le l\le2000$.

\begin{figure}
 \centering
\includegraphics[scale=.23]{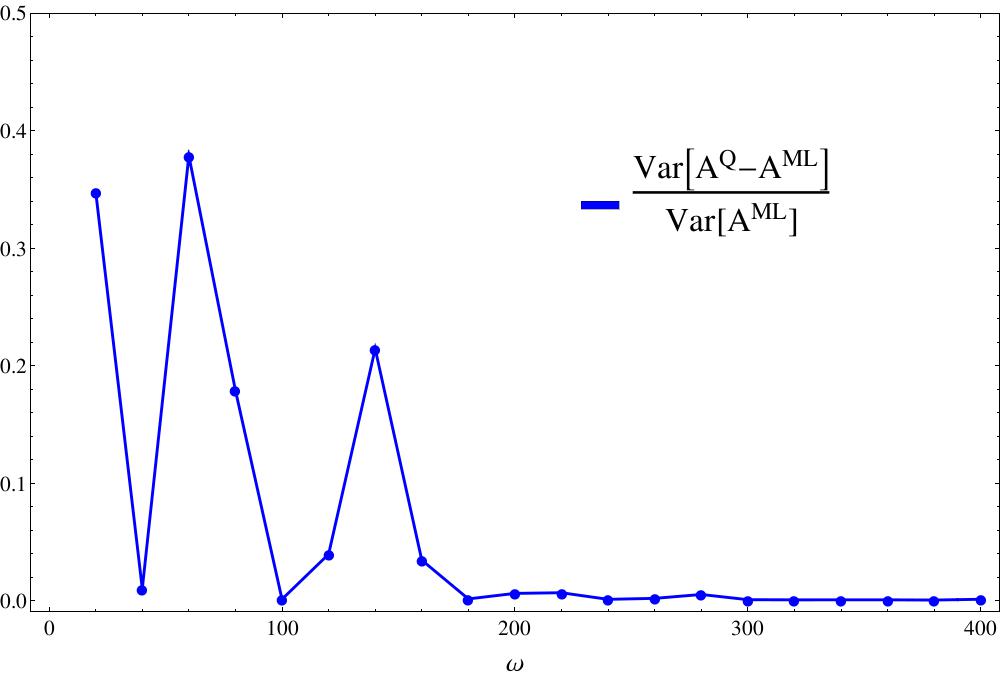}
\caption{Ratio of the variance of the difference between the full ML estimate $\hat{A}^{ML}_P$ and an approximation based on first finding the best-fit cosmological parameters in the absence of features and then using a quadratic estimator $\hat{A}^Q_P$ to obtain the amplitudes. We obtain the variance at various frequencies $\omega$ for assuming a phase $\phi_P=0$.}
\label{fig:MLvalid}
\end{figure}

In linear theory, in particular ignoring lensing, the observed CMB power spectrum given the six $\Lambda$CDM parameters $p_i$ and a feature model with a certain $\omega$, $\phi$ is given by
\begin{equation}
C_l(p_i, A)=C_l^{\Lambda\text{CDM}}(p_i)+A_P\,\delta C_l(p_i,\omega, \phi)
\end{equation}
Determining the ML estimate for the amplitude $\hat{A}_P\equiv\hat{A}_P^{\text{ML}}$ in principle requires all parameters to be varied simultaneously. I.e. for each point of the $\omega$-$\phi$ grid one has to vary the amplitude as well as the six $\Lambda$CDM parameters. This is a computationally intensive task. Elaborate searches for oscillations in the power spectrum have been performed elsewhere (see for example \cite{Meerburg:2011WMAP7Constraints,Peiris:2013ConstMonoInf,Flauger:2009MonoInf,Meerburg:2013SearchOscP1,Meerburg:2013SearchOscP2,Meerburg:2014SearchOsc}). Rather than trying to obtain the best-fit amplitude to very high precision we implement a pipeline that should provide reliable results for a given map with reasonable computational effort and sufficient accuracy to be able to make statements about statistical dependencies of measurements. It needs to be fast enough to allow the processing of a large number of CMB realisations. We neglect the effect of varying the cosmological parameters on the perturbation to the observed power spectrum $\delta C_l$. Instead, we precompute it employing CAMB \cite{Lewis:CAMB} with sufficiently high precision settings for each $\omega$, $\phi$ using the Planck best-fit $\Lambda$CDM parameters. Using the precomputed oscillatory component means that we can run CAMB with lower precision settings for finding the ML point as the remaining $\Lambda$CDM varies relatively slowly. Varying all parameters for each point on the grid still remains a challenging task and we adopt a further simplification. At the lowest frequencies where oscillatory features are to some extent degenerate with changes in the cosmological parameters a joint ML estimate is certainly the only reliable option. However, one expects that the ML amplitude $\hat{A}_P$ should decouple from the cosmological parameters at higher frequency in the sense that we can first search for the best-fit $\Lambda$CDM model setting the amplitude to zero and then obtain the amplitude keeping the cosmological parameters fixed. The second option has the advantage that once we found the best-fit $\Lambda$CDM model, the best-fit amplitudes can be found as a simple quadratic estimate%
\footnote{We introduce a redundant factor of 2 in the definition of the quadratic estimator here for consistency with the standard optimal power spectrum estimator and the optimal bispectrum estimator later on. This definition implies $\<\hat{A}^2_P\>=2!/N_P$ in line with $\<\hat{A}_B^2\>=3!/N_B$ for the bispectrum.}%
\begin{eqnarray}
\hat{A}^Q_P&=&\frac{2}{N}\delta C_{l_1}\Delta^{-1}_{l_1l_2}\left(\hat{C}_{l_2}-C^{\Lambda\text{CDM,ML}}_{l_2}\right)\\
N_P&=&2\delta C_{l_1}\Delta^{-1}_{l_1l_2}\delta C_{l_2}
\end{eqnarray}
for each point $(\omega,\phi)$. This reduces the computational effort tremendously. Figure \ref{fig:MLvalid} compares the amplitudes $\hat{A}^Q_P$ and those obtained as a ML estimate by varying cosmological parameters and the amplitude simultaneously $\hat{A}^{ML}_P$. We plot the ratio of variance of the difference and the variance of the joint ML estimates for various frequencies and $\phi_P=0$.
One can clearly see that the degeneracies between cosmological parameters and the feature models at low frequencies lead to differences in these amplitude estimates. This is true in particular at $\omega\sim 70$ and $\omega\sim 140$ (peaks of decreasing height also appear at $\omega\sim 210, 280, \ldots$ but are barely visible in this plot). The oscillation in the power spectrum due to the acoustic peaks is determined by the comoving sound horizon at last scattering $r_S$ and resembles a primordial oscillation with $\omega\sim 140$ (roughly the value of $r_S$ in Mpc) or equivalently%
\footnote{Projection relates a mode at last scattering with comoving scale $k$ to an $l$ scale roughly given by $l\sim\Delta\eta k$ where $\Delta\eta$ is the comoving distance to last scattering. Hence, a feature in the power spectrum with frequency $\omega$ gives rise to oscillations in $l$ with wavelengths $\lambda_l\sim\pi\Delta\eta/\omega$. With $\Delta\eta\sim1.4\times 10^4\text{Mpc}$ we see that $\omega\sim 140$ produces oscillations with an approximate wavelength of $\lambda_l\sim 300$ that mimics the acoustic peak structure.} %
an oscillation in $l$ with $\lambda_l\sim 300$.
Thus this is an intuitive result reflecting that changes in the $\Lambda$CDM cosmology can mimic the effect of a feature particularly well when $2\,\omega\sim n\,r_s$ for some integer $n$ because such a feature either increases (or decreases) the height of the peaks or changes the relative height of neighbouring peaks. Beyond $\omega\sim 140$ any differences become very small.

We conclude that first finding the best-fit $\Lambda$CDM cosmology and then using a quadratic estimate to find the amplitudes is a relatively accurate method to determine the ML amplitudes. It certainly offers enough accuracy for the present study so that we adopt it as our power spectrum pipeline.

Note that using these simplifications there is a close link between the amplitudes $\hat{A}_P$ and the corresponding likelihood improvement that is often studied to search for oscillatory features. After the best-fit $\Lambda$CDM model has been found, the improvement in the likelihood from varying the amplitude alone is simply given by $-2\Delta\log{\cal{L}}=-N/2\,(\hat{A}_P^{\text{ML}})^2$.

Summing up, for each map we first find the best-fit cosmological parameters using the optimisation routine BOBYQA \cite{BOBYQA} and CAMB. With appropriate settings we find that the routine is able to determine the ML point of the likelihood to sufficient accuracy with $\mathcal{O}(10^2)$ calls of CAMB, making this procedure fast enough to be performed on a large number of CMB realisations. We then determine the best-fit amplitudes $\hat{A}^{ML}$ using a quadratic estimator as outlined above.

\subsection{Bispectrum pipeline}
To constrain feature models via the bispectrum we use the modal polynomial pipeline that was used in the 2013 Planck analysis \cite{Ade:2013CosmicStrings, Ade:2013NonGaussianity, Ade:2013IsotropyStatistics, Ade:2013ISW}. This is an implementation of the standard optimal bispectrum estimator. The optimal estimator for the bispectrum amplitude of a feature model $\hat{A}_B$ reads
\begin{widetext}
\begin{equation}
\hat{A}_B=\frac{1}{N_B}\sum_{\ell^{\phantom{'}}_i \ell'_i m^{\phantom{'}}_i m'_i}  \curl{G}^{\ell^{\phantom{'}}_1 \ell^{\phantom{'}}_2 \ell^{\phantom{'}}_3}_{m^{\phantom{'}}_1 m^{\phantom{'}}_2 m^{\phantom{'}}_3} b_{\ell^{\phantom{'}}_1 \ell^{\phantom{'}}_2 \ell^{\phantom{'}}_3} (C^{-1})_{\ell^{\phantom{'}}_1\ell'_1m^{\phantom{'}}_1m'_1} (C^{-1})_{\ell^{\phantom{'}}_2\ell'_2m^{\phantom{'}}_2m'_2} (C^{-1})_{\ell^{\phantom{'}}_3 \ell'_3 m^{\phantom{'}}_3 m'_3} \left(a_{\ell'_1 m'_1} a_{\ell'_2 m'_2} a_{\ell'_3 m'_3} - 3\<a_{\ell'_1 m'_1} a_{\ell'_2 m'_2}\> a_{\ell'_3 m'_3}\right)\,,
\end{equation}
\end{widetext}
where $b$ is the theoretical bispectrum of the feature model defined by
\begin{equation}
\< a_{\ell_1 m_1} a_{\ell_2 m_2} a_{\ell_3 m_3} \> = B^{l_1l_2l_3}_{m_1m_2m_3}=\curl{G}^{\ell_1 \ell_2 \ell_3}_{m_1 m_2 m_3} b_{\ell_1 \ell_2 \ell_3}\,,
\end{equation}
and $\curl{G}$ is the Gaunt integral, which is the projection of the angular part of the primordial delta function
\begin{widetext}
\begin{eqnarray}
\nonumber\curl{G}^{\ell_1 \ell_2 \ell_3}_{m_1 m_2 m_3}&=&\int d\Omega_{\un} Y_{\ell_1 m_1 }(\un) Y_{\ell_2 m_2 }(\un) Y_{\ell_3 m_3 }(\un)=\left(\begin{array}{ccc}\ell_1 & \ell_2 & \ell_3 \\ m_1 & m_2 & m_3  \end{array}\right) h_{\ell_1 \ell_2 \ell_3}\,, \\
h_{\ell_1 \ell_2 \ell_3}&=&\sqrt{\frac{(2\ell_1+1)(2\ell_2+1)(2\ell_3+1)}{4\pi}}\left(\begin{array}{ccc}\ell_1 & \ell_2 & \ell_3 \\ 0 & 0 & 0  \end{array}\right)\,.
\end{eqnarray}
The normalisation of the estimator is
\begin{equation}\label{eq:norm}
N_B \equiv \sum_{\ell^{\phantom{'}}_i\ell'_i}\curl{G}^{\ell^{\phantom{'}}_1 \ell^{\phantom{'}}_2 \ell^{\phantom{'}}_3}_{m^{\phantom{'}}_1 m^{\phantom{'}}_2 m^{\phantom{'}}_3} b_{\ell^{\phantom{'}}_1 \ell^{\phantom{'}}_2 \ell^{\phantom{'}}_3} (C^{-1})_{\ell^{\phantom{'}}_1\ell'_1m^{\phantom{'}}_1m'_1} (C^{-1})_{\ell^{\phantom{'}}_2\ell'_2m^{\phantom{'}}_2m'_2} (C^{-1})_{\ell^{\phantom{'}}_3\ell'_3m^{\phantom{'}}_3m'_3} \curl{G}^{\ell'_1 \ell'_2 \ell'_3}_{m'_1 m'_2 m'_3} b_{\ell'_1 \ell'_2 \ell'_3}\,.
\end{equation}
\end{widetext}
The pipeline employs separable basis functions to dramatically simplify the calculation and allow us to constrain all models simultaneously. The approach was first described in \cite{Fergusson:PrimordialNonGaussianity} and a fully realised version was first implemented in \cite{Fergusson:BispectrumEstimationI}, when it was applied to WMAP data. It was recently extended to polarisation in preparation for the next round of Planck papers in \cite{Fergusson:Efficient}, which also included many other small advances. It is the temperature only version of this pipeline that was used here and we refer the interested reader to the previous reference for a full description of the method.  This pipeline has proved very efficient for scanning large parameter spaces quickly for a broad selection of models and was extensively validated as part of the Planck analysis. This version has undergone significant optimisation since then and so we were able to increase the number of basis functions from 601 to 2001, doubling our frequency coverage.

\section{Results}
\label{sec:Results}
The pipelines described in Sec.~\ref{sec:Methods} produce amplitudes $\hat{A}_i$ for each point on the $\omega$-$\phi$ grid. Here, $i=P$ for the power spectrum and $i=B$ for the bispectrum. We use a grid covering the range $\omega=10-600$ in steps of $\Delta\omega=10$ and 10 steps in phase with $\phi=0, \pi/10, \ldots, 9/10\pi$. Due to correlations between nearby frequencies increasing the resolution beyond $\Delta\omega=10$ has little benefit (at least given an $l$-range $50\le l\le 2000$) as will be discussed below.   
The central question we are trying to answer is whether or not there are any significant statistical dependencies between the amplitudes $\hat{A}_P(\omega_P,\phi_P)$ measured in the power spectrum and the amplitudes $\hat{A}_B(\omega_B,\phi_B)$ measured in the bispectrum. We know that the amplitudes must be uncorrelated, $\text{Corr}(\hat{A}_P,\hat{A}_B)=0$. To investigate more complicated dependencies we use two different measures. The standard correlation coefficient of the absolute values of the amplitudes $\text{Corr}(|\hat{A}_P|,|\hat{A}_B|)$ and the distance correlation coefficient $\text{dCorr}(\bar{A}_P,\bar{A}_B)$.

\subsection{Correlation between amplitude measurements}
\begin{figure*}
 \centering
\includegraphics[scale=.5]{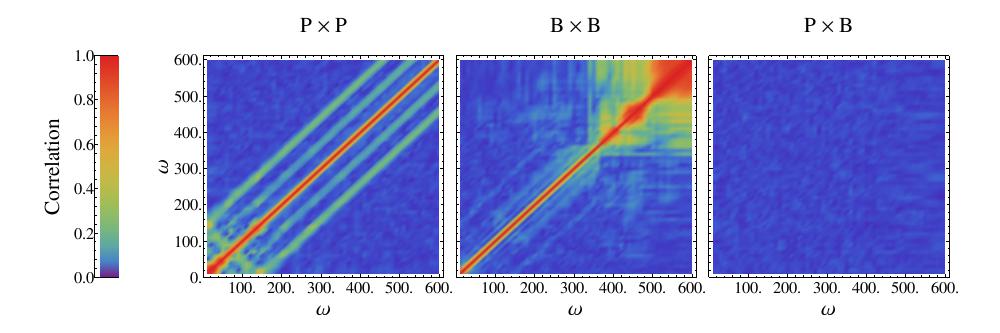}
\caption{Correlation matrices obtained from 10000 noiseless full-sky CMB realisations. For each frequency combination we plot the largest absolute value of the correlation $\text{Corr}(|\hat{A}_i|,|\hat{A}_j|)$ found amongst all the possible combinations of phases.}
\label{fig:Corr10k}
\end{figure*}

Given a set of $N$ simulations we estimate the correlation between the absolute values of different amplitudes as
\begin{equation}
\text{Corr}(|\hat{A}_i|,|\hat{A}_j|)=\frac{\sum_n\left(|\hat{A}_i^n|-\mu_i\right)\left(|\hat{A}_j^n|-\mu_j\right)}{\sqrt{\sum_n\left(|\hat{A}_i^n|-\mu_i\right)^2\sum_n\left(|\hat{A}_j^n|-\mu_j\right)^2}}
\end{equation}
where $\hat{A}_i^n$ refer to the amplitudes obtained from simulation $n$ and $\mu_i=(\sum |\hat{A}_i^n|)/N$ is simply the mean of the measurements. Note that $\mu_i\ne0$ because we are studying the absolute values of the amplitudes. We obtained the correlation matrix for both sets of simulations. Figure \ref{fig:Corr10k} shows the results in the case of the 10000 simple noiseless full-sky simulations.
To visualise the results better, for each pair $\omega_P$, $\omega_B$ we plot the maximum magnitude correlation found in the sample for any choice of $\phi_P$, $\phi_B$. We see that there is some off-diagonal structure at the ${\cal{O}}(10\%)$ in the power spectrum-power spectrum (PP) correlation plots. This is to be expected as there is clearly some degeneracy between different oscillatory models. The correlation shows up as narrow stripes parallel to the diagonal with the most pronounced stripe at a distance of about $\Delta\omega\sim 140$. This is approximately the frequency of the acoustic peaks in the power spectrum as we discussed in Sec.~\ref{sec:Methods}. We can understand this by recalling that $\sin(\omega_1 k+\phi_1)\sin(\omega_2 k+\phi_2)= \cos((\omega_1-\omega_2)k+(\phi_1-\phi_2))+\cos((\omega_1+\omega_2) k+(\phi_1+\phi_2))$, so that if $\omega_1-\omega_2\sim 140$ there is some resonance with the acoustic peaks that gives rise to excess correlation for appropriate choices of the relative phases of the features. The bispectrum-bispectrum (BB) correlations are very nearly diagonal up to $\omega\sim 400$ implying that different feature models are almost independent. The off-diagonal structure visible at high $\omega$ is due to insufficient convergence of the modal expansions of the corresponding shapes using 2000 modes. The fact that modulations close in frequency are strongly correlated in both the power spectrum and the bispectrum so that the diagonals have finite width suggests that there is an effective stepwidth $\Delta\omega_{\text{eff}}\gtrsim10$. This will be of importance later on when we study the statistics of feature model surveys. 

For a joint analysis it is important to study correlations between the two measurements. Crucially, we see that there is no sign of any power spectrum-bispectrum (PB) correlation at the  1\% level that we can resolve with 10000 samples.

\begin{figure*}
 \centering
\includegraphics[scale=.23]{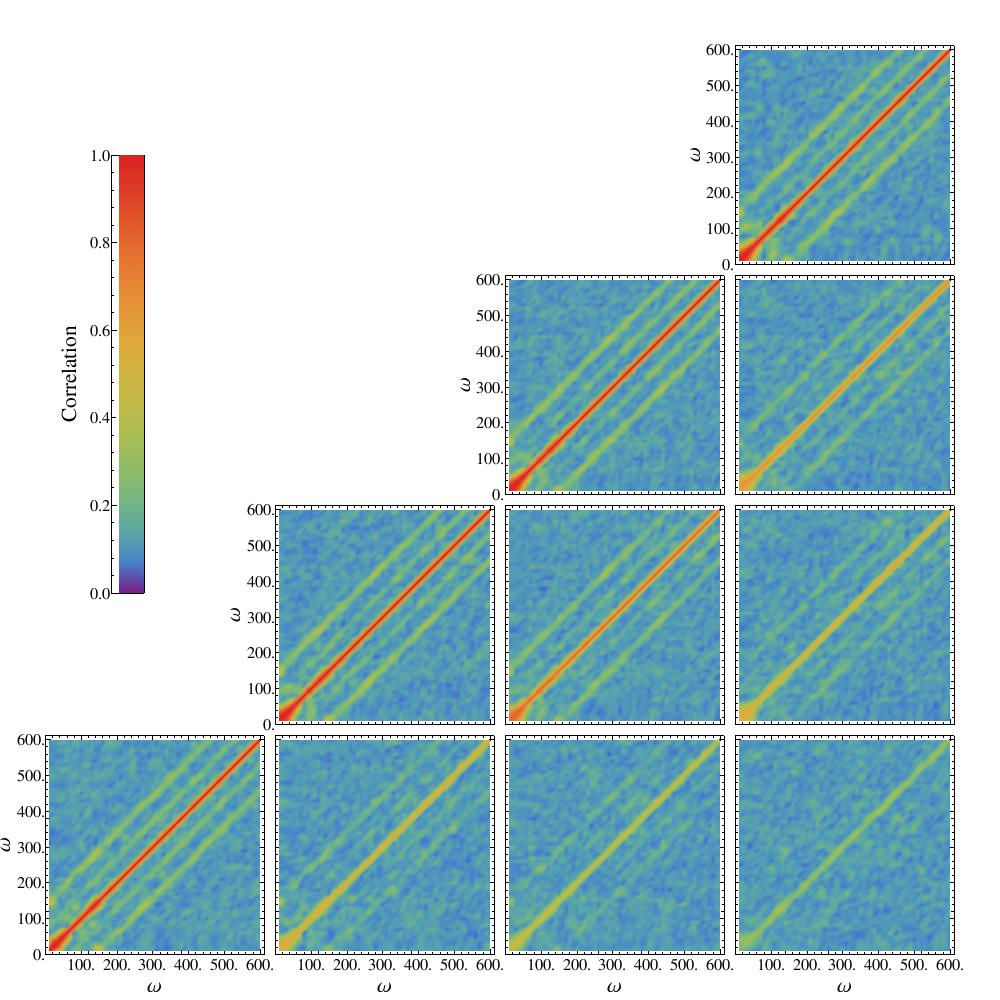}
\includegraphics[scale=.23]{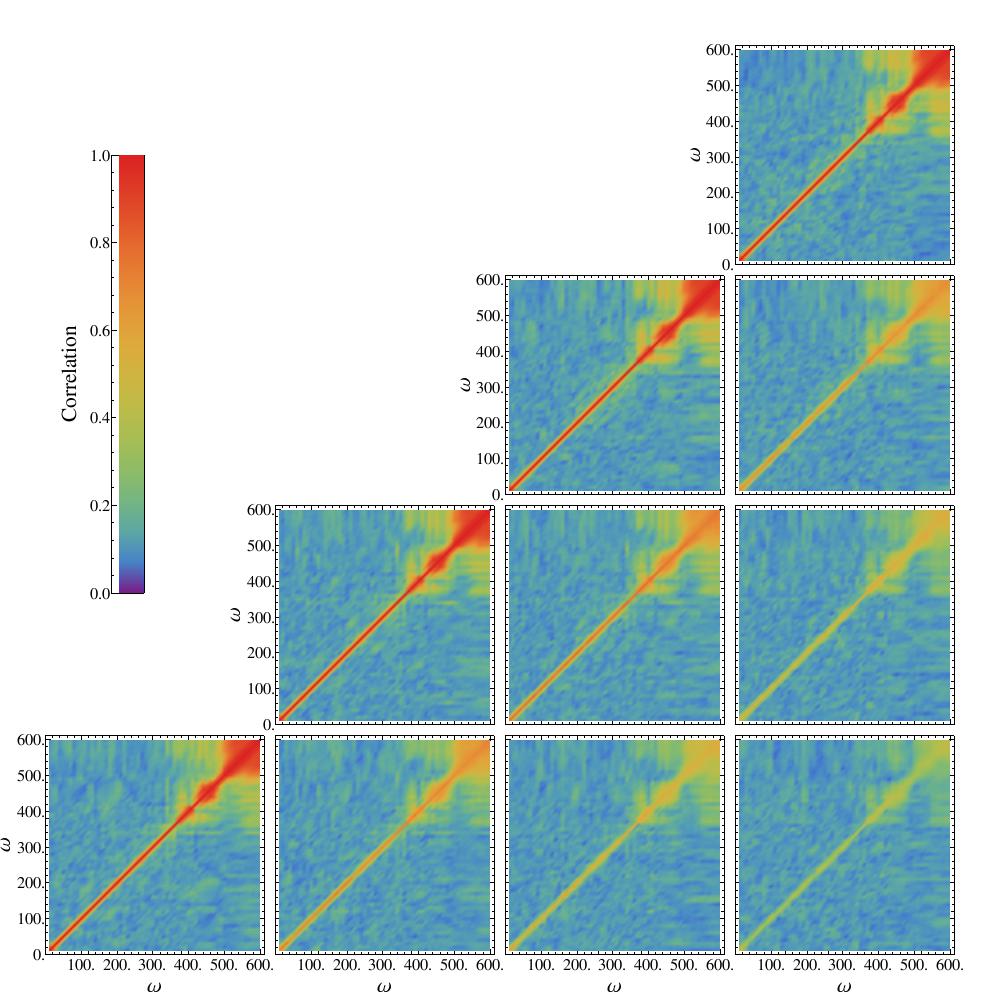}
\includegraphics[scale=.23]{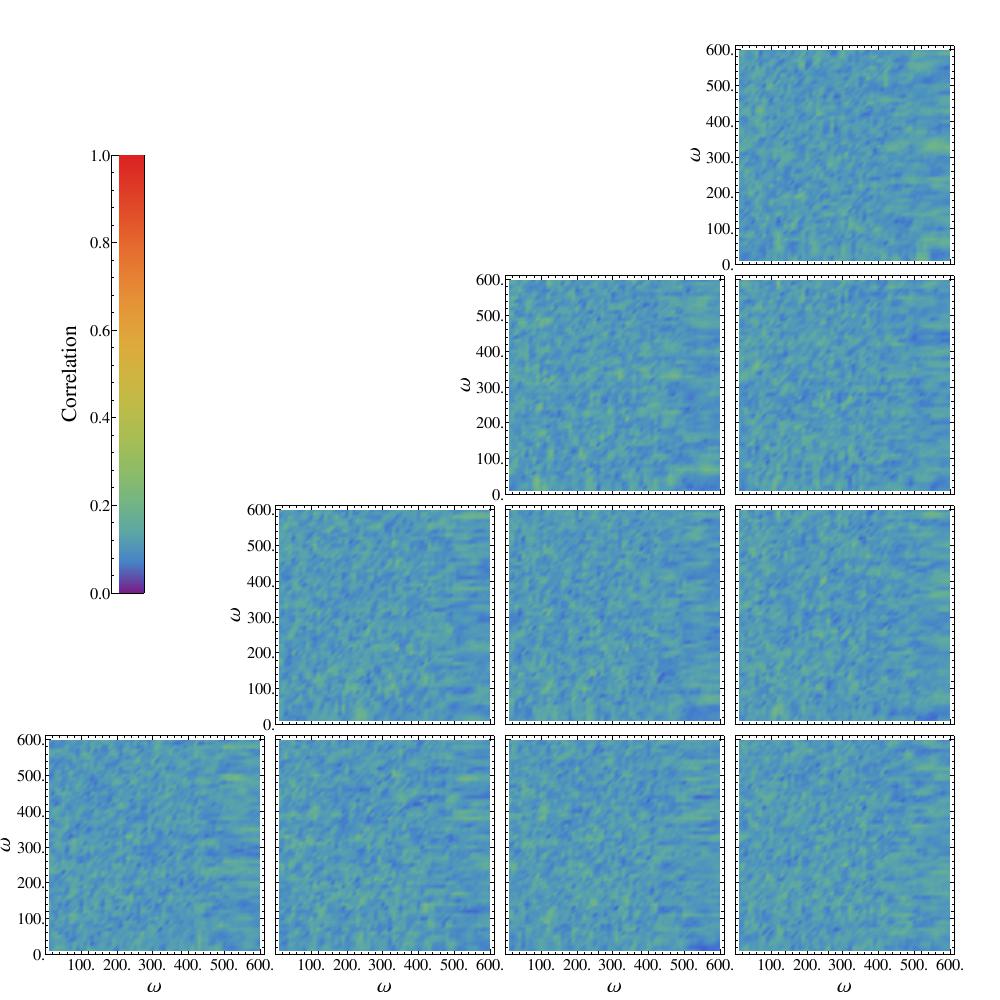}
\caption{Correlation matrices obtained from 400 CMB realisations with anisotropic noise. The results are shown for complete sky coverage as well as using the U73, MU56 and MU38 masks. As in Fig.~\ref{fig:Corr10k} we plot the largest absolute value of $\text{Corr}(|\hat{A}_i|,|\hat{A}_j|)$ for a given $\omega$ combination found amongst all the possible combinations of phases.}
\label{fig:Corr400}
\end{figure*}

Knowing that there is no significant correlation between the measurements in the simplified case of full-sky noiseless CMB realisations a natural question to ask is whether features of real CMB experiments such as complicated instrumental noise or incomplete sky coverage might change this conclusion. We investigated this question using the 400 realistic simulations described in Sec.~\ref{sec:Methods}. Figure \ref{fig:Corr400} shows the corresponding correlation plots. 
The MC noise in these plots is clearly larger, allowing a reliable detection of correlations down to the 5\% level. Barring the higher noise level, for each mask the plots agree extremely well with the correlation plots in the simplified case suggesting that the inclusion of anisotropic noise and masking of parts of the sky has little effect. The PP and BB correlations between measurements using different masks show less correlation when the difference in sky fraction is larger. This is intuitive given that the amount of data that is included in one of the measurements but not in the other increases.   

\subsection{Distance correlation between amplitude measurements}
\begin{figure*}
 \centering
\includegraphics[scale=.5]{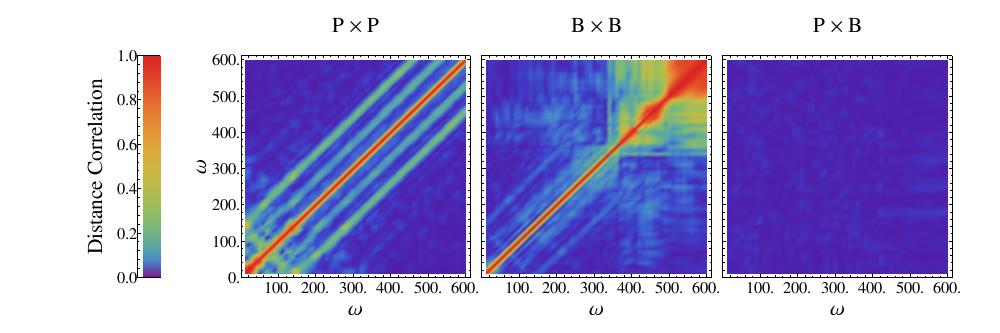}
\caption{Distance correlation matrices obtained from 1000 noiseless full-sky CMB realisations. For each combination of frequencies $\omega$ the largest absolute value of $\text{dCorr}(\hat{A}_i,\hat{A}_j)$ for any combination of phases is plotted.}
\label{fig:dCorr1k}
\end{figure*}

In the previous section we showed that if there exist any correlations between the absolute values of the amplitudes $\hat{A}_P$ and $\hat{A}_B$ they must be at the $\lesssim 1\%$ level, the level at which we can hope to detect correlations on the basis of 10000 samples. While the correlation matrix is a useful and familiar tool to detect statistical dependencies, a vanishing correlation coefficient does not imply statistical independence. To investigate whether there are any detectable dependencies in the sets of simulations that do not cause linear correlation we also calculated the distance correlation matrix of the amplitudes. Distance correlation has the property that it vanishes if and only if the random variables are truly independent. It is defined as
\begin{equation}
\text{dCorr}(\hat{A}_i,\hat{A}_j)=\frac{\sum_{k,l}\Delta_{kl}^i\Delta_{kl}^j}{\sqrt{\sum_{k,l}(\Delta_{kl}^i)^2\sum_{k,l}(\Delta_{kl}^j)^2}}
\end{equation}
where
\begin{eqnarray}
\Delta_{kl}^i&=&\delta_{kl}^i-\frac{\sum_m\delta_{km}^i}{N}-\frac{\sum_m\delta_{ml}^i}{N}+\frac{\sum_{mn}\delta_{mn}^i}{N^2}\\
\delta_{kl}^i&=&\left|\hat{A}_i^k-\hat{A}_i^l\right|
\end{eqnarray}
As this expression is relatively tedious to calculate owing to the fact that the distance between all pairs of amplitudes from different samples enters this expression we evaluated it only on the first 1000 samples of full sky noiseless CMB realisations. The results are shown in Fig.~\ref{fig:dCorr1k}.
The result is qualitatively very similar to Fig.~\ref{fig:Corr10k}. The main differences are a somewhat smaller noise level, a consequence of summing over $\mathcal{O}(10^6)$ pairs of samples, and slightly different correlation coefficients. Slight differences in the correlation coefficients are to be expected because there is no direct correspondence between the precise values of correlation and distance correlation except for the case of statistically independent variables in which both give 0 and the case of fully linearly dependent variables in which both are unity.
\begin{figure*}
 \centering
\includegraphics[scale=.23]{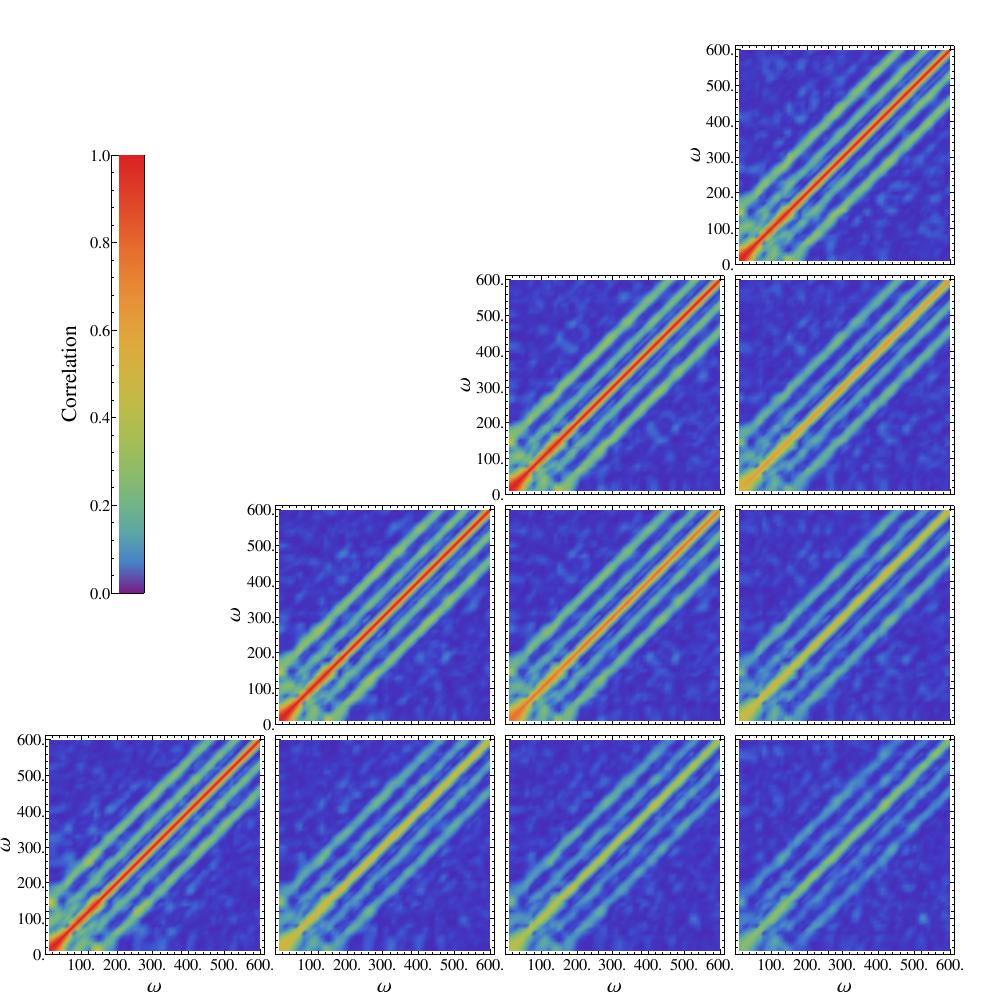}
\includegraphics[scale=.23]{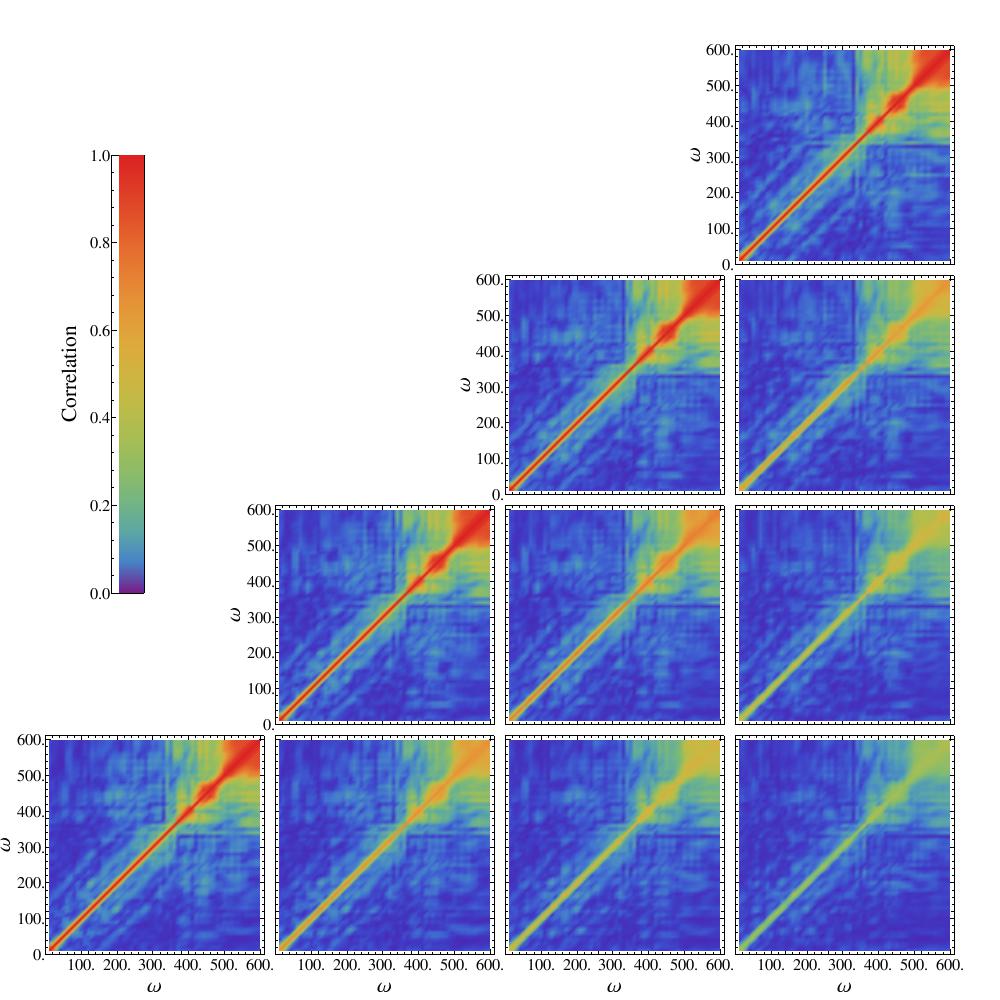}
\includegraphics[scale=.23]{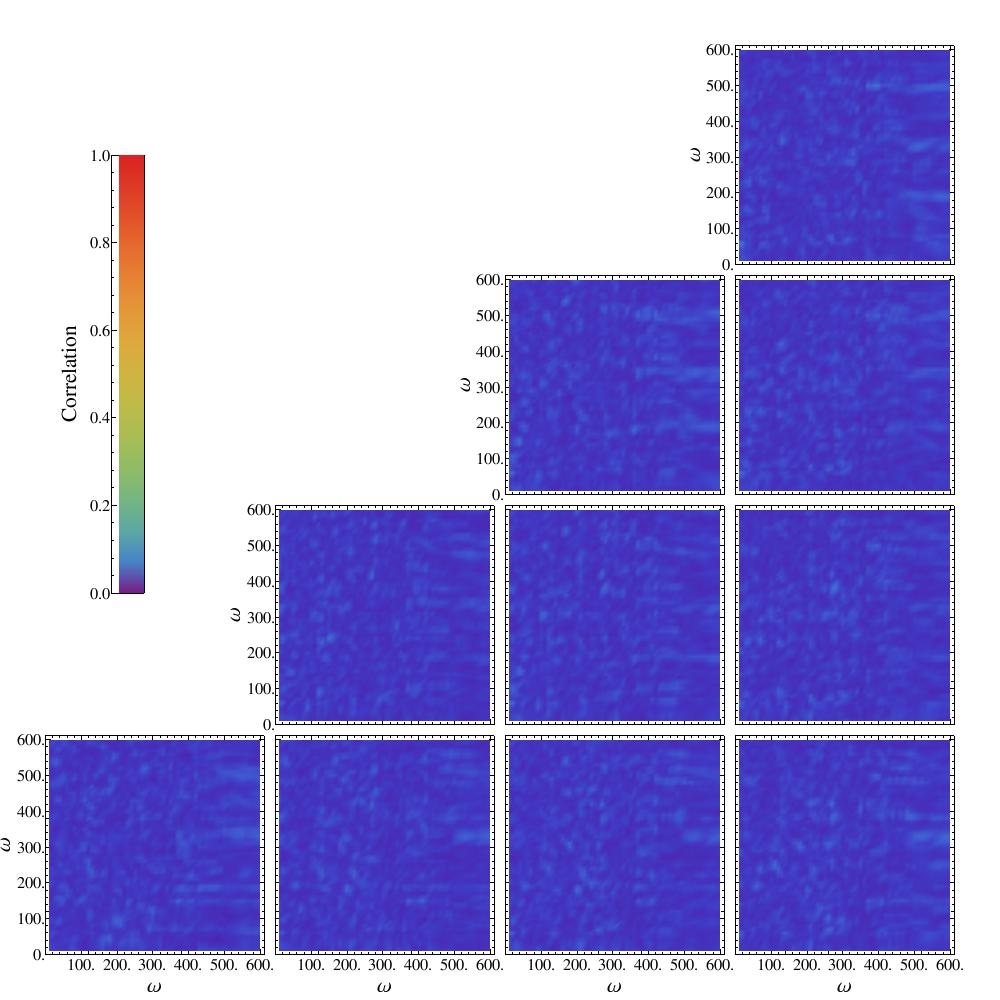}
\caption{Distance correlation matrices obtained from 400 CMB realisations with anisotropic noise. The results are shown for complete sky coverage as well as using the U73, MU56 and MU38 masks. Again we plot the largest absolute value of $\text{dCorr}(\hat{A}_i,\hat{A}_j)$ for a given combination of $\omega$.}
\label{fig:dCorr400}
\end{figure*}
Figure \ref{fig:dCorr400} shows the distance correlation matrices obtained from the 400 realistic simulations. Again, the plot is qualitatively very similar to the corresponding correlation plots in Fig.~\ref{fig:Corr400} and confirms the conclusions drawn at the end of the previous section.

\subsection{Analytic estimate}
Having found no detectable dependencies between amplitude measurements it is desirable to have an analytic understanding of this fact. Heuristically we could make the following point. While the amplitudes $\hat{A}_P$ and $\hat{A}_B$ might not be jointly Gaussian distributed for finite sample sizes, we could argue that as we increase $l_{\text{max}}$, according to the general expectation from the multidimensional central limit theorem (CLT), we expect the amplitudes to approach a joint distribution close to Gaussian. This is clearly not a proof as strictly speaking the requirements of the CLT are not met. Nonetheless, if we assume that the joint distribution approaches a Gaussian, uncorrelatedness implies statistical independence. As the estimators for the power spectrum and bispectrum amplitude are always uncorrelated, this argument leads us to conclude that the power spectrum and bispectrum amplitudes should be asymptotically independent.

To gain further insight we are going to derive an estimate for the degree of dependence in the simple isotropic Gaussian case. Rather than the absolute values we study the variance of the square of the amplitudes
\begin{equation}
\text{Corr}\left[\hat{A}_P^2,\hat{A}_B^2\right]=\frac{\text{Cov}\left[\hat{A}_P^2,\hat{A}_B^2\right]}{\left(\text{Var}\left[\hat{A}_P^2\right]\text{Var}\left[\hat{A}_B^2\right]\right)^{\frac{1}{2}}}
\end{equation}
For independent amplitudes we obviously have $\text{Corr}\left[\hat{A}_P^2,\hat{A}_B^2\right]=0$ but we are faced with a ten-point function so we cannot conclude that it must vanish just because the CMB is Gaussian. However, the only possible contributions to the variance are either of the form $\delta_1$ or $\delta_2$ where
\begin{widetext}
\begin{eqnarray}
\delta_1&=&\frac{1}{(N_P\,N_B)^2}B_{l_1m_1l_2m_2l_3m_3}C^{-1}_{l_1}C^{-1}_{l_2}C^{-1}_{l_3}B_{l_1m_1l_2m_2l_3m_3}\frac{\delta C_{l_2}}{C_{l_2}}\frac{\delta C_{l_3}}{C_{l_3}}\\
\delta_2&=&\frac{1}{(N_P\,N_B)^2}B_{l_1m_1l_2m_2l_3m_3}C^{-1}_{l_1}C^{-1}_{l_2}C^{-1}_{l_3}B_{l_1m_1l_2m_2l_3m_3}\left(\frac{\delta C_{l_3}}{C_{l_3}}\right)^2
\end{eqnarray}
\end{widetext}  
with some combinatorical prefactors. Now in general we expect $\delta C_{l}/C_{l}\sim N_P^{1/2}/l^2_{\text{max}}$. Furthermore we assume $\text{Var}\left[\hat{A}_P^2\right]\sim\text{Var}\left[\hat{A}_P\right]^2\sim N_P^{-2}$ and similarly for the bispectrum so that as an order of magnitude estimate we have
\begin{eqnarray}\nonumber
&\quad&\text{Corr}\left[\hat{A}_P^2,\hat{A}_B^2\right]\\
&\sim&\frac{(N^2_P\,N^2_B)^{\frac{1}{2}}}{N^2_P\,N^2_B} \underbrace{(BC^{-1}C^{-1}C^{-1}B)}_{=N_B} \frac{N_P}{l^2_{\text{max}}}\sim\frac{1}{l^2_{\text{max}}}
\end{eqnarray}
Thus we expect $\delta\sim l^{-2}_{\text{max}}$.

This argument strictly holds only for the square of the amplitudes but it strongly suggests that we expect any dependencies between the amplitudes themselves to be suppressed by a factor $l_{\text{max}}^{-1}$ and in particular $\text{Corr}\left[|\hat{A}_P|,|\hat{A}_B|\right]\sim\text{dCorr}\left[A_P,A_B\right]=\mathcal{O}(l_{\text{max}}^{-1})$. Hence any correlations should be of the order $10^{-3}$ and thus too small to be detected with $10^4$ samples and certainly too small to be relevant for the detection of feature models. This argument should add further credibility to the results presented above.

\section{Combined statistics}
\label{sec:CombSearch}
\subsection{Combined search for feature models}
\label{subsec:CombStat}
Having shown that we can expect largely independent constraints on feature models from the power spectrum and bispectrum the natural question is how we design a combined search for feature models. One route would be to pursue a combined likelihood analysis. We write the contribution of the feature model to the two- and three-point correlator as
\begin{eqnarray}\nonumber
\<a_{l_1m_1}a_{l_1m_1}\>&=&C_{l_1m_1l_2m_2}\\
&=&C^0_{l_1m_1l_2m_2}+\delta C_{l_1m_1l_2m_2}\\
\<a_{l_1m_1}a_{l_2m_2}a_{l_3m_3}\>&=&B^{l_1l_2l_3}_{m_1m_2m_3}
\end{eqnarray}
where $C^0$ is the two-point correlator in the absence of any feature model. We can write down a formal expression for the full non-Gaussian likelihood using the Edgeworth expansion in the connected $n$-point functions (see for example \cite{Babich:OptEst, Taylor:Edgeworth} and references within for details)
\begin{widetext}
\begin{eqnarray}
\mathcal{P}\left(a_{lm}\right)&=&\exp\left(\sum_{n\ge3}\frac{\imath^n}{n!}\<a_{l_1m_1}\ldots a_{l_nm_n}\>_c\frac{\partial}{\partial a_{l_1m_1}}\ldots\frac{\partial}{\partial a_{l_nm_n}}\right)\frac{\exp{\left(-\frac{1}{2}C^{-1}_{l_1m_1l_2m_2}a_{l_1m_1}a_{l_2m_2}\right)}}{\sqrt{2\pi\det{(C)}}}\\
&=&\exp\left(1-\frac{1}{3!}B^{l_1l_2l_3}_{m_1m_2m_3}\frac{\partial}{\partial a_{l_1m_1}}\frac{\partial}{\partial a_{l_1m_1}}\frac{\partial}{\partial a_{l_1m_1}}+\mathcal{O}(f_{NL}^2)\right)\frac{\exp{\left(-\frac{1}{2}C^{-1}_{l_1m_1l_2m_2}a_{l_1m_1}a_{l_2m_2}\right)}}{\sqrt{2\pi\det{(C)}}}
\end{eqnarray}
\end{widetext}
This expression needs to be expanded to high enough order in the characteristic amplitude of deviations from Gaussianity denoted by $f_{NL}$ to guarantee sufficient accuracy. Note that it is not enough to know the three-point function of the feature model for an accurate likelihood. We need knowledge of the contributions to higher order connected correlation functions as well.

Without calculating the full likelihood there is another natural statistic to consider. Assuming the feature model has an overall amplitude $A$ as a parameter, i.e. $\<aa\>=C^0+A\delta C$, $\<aaa\>=A B$ where $\delta C$ and $B$ depend on other parameters of the feature model that are scanned over, we can easily construct the optimal unbiased estimator for said amplitude at $A\approx 0$. To do so we assume that $A$ is not degenerate with other cosmological parameters influencing $C^0$ so that we can estimate these parameters separately in a first step and take $C^0$ to be a fixed fiducial covariance. As we have seen in Sec.~\ref{sec:Methods} this is the case except at low $\omega$. The optimal amplitude estimator is then simply given by
\begin{eqnarray}
\hat{A}&=&\frac{1}{N}\left(\frac{\hat{A}_P}{V_P}+\frac{\hat{A}_B}{V_B}\right)\\
N&=&\frac{1}{V_P}+\frac{1}{V_B}
\end{eqnarray}
Here $\hat{A}_P$, $\hat{A}_B$ are the standard amplitude estimators discussed above for the power spectrum modulation $\delta C$ and the bispectrum $B$ with variances $V_P=\<\hat{A}_P^2\>=2/N_P$ and $V_B=\<\hat{A}_B^2\>=3!/N_B$. The optimality of this estimator can be simply seen from the Edgeworth expansion above just like it is argued in \cite{Babich:OptEst} for the case of the bispectrum-only estimator. Expanding to first order in $A$ we obtain
\begin{equation}
\frac{\partial\log{\mathcal{P}}}{\partial A}\Big|_{A=0}=\frac{N_P}{2}\hat{A}_P+\frac{N_B}{3!}\hat{A}_B=\frac{\hat{A}_P}{V_P}+\frac{\hat{A}_B}{V_B}
\end{equation}
so the Fisher information is
\begin{equation}
\mathcal{F}_A=\<\left(\frac{\partial\log{\mathcal{P}}}{\partial A}\right)^2\>\Big|_{A=0}=\<\left(\frac{\hat{A}_P}{V_P}+\frac{\hat{A}_B}{V_B}\right)^2\>\Big|_{A=0}
\end{equation}
At $A=0$ we have $\<\hat{A}_P\hat{A}_B\>=0$ because this is effectively a five-point function that must vanish in the case of a Gaussian distribution so that $\mathcal{F}_A=N$ and
\begin{equation}
\<\hat{A}^2\>=\frac{1}{N}=\frac{1}{\mathcal{F}_A}
\end{equation}
and $\hat{A}$ indeed saturates the Cramer-Rao bound for unbiased estimators.

As we are considering models that have individual amplitude parameters for the power spectrum and bispectrum $A_P$ and $A_B$, they can be easily parametrised by an overall amplitude $A$. If we define a relative amplitude $r:=A_B/A_P$, then for a given $r$ the optimal amplitude estimator is
\begin{equation}
\hat{A}=\frac{\hat{A}_P/V_P+r\hat{A}_B/V_B}{V_P^{-1}+r^2V_B^{-1}}
\end{equation}
where $\hat{A}_P$, $\hat{A}_B$, $V_P$ and $V_B$ now refer to the amplitude estimators for the bare sine modulations in the power spectrum and bispectrum.

\subsection{Quantifying the look-elsewhere effect in feature model surveys}
After the discussion in the last section it is of interest to answer the question whether there is any hope of finding significant evidence for feature models in a combined survey when individual searches both produced no convincing evidence. Whether one studies likelihood improvements or optimal amplitude estimates, a sensible question to judge whether one should get excited about the results or not is to compare the findings to what one would expect from simply fitting a random Gaussian realisation. As feature model surveys typically scan a large number of models there is a significant look-elsewhere effect involved, i.e. we expect to see naively significant results simply because we tried many different models. So far surveys for the power spectrum and bispectrum have not produced any results that exceed the significances one would expect from a featureless Gaussian realisation. A central goal of this section is to show that even with additional tunable parameters such as the relative amplitude and the relative phase of feature models that are introduced in the generic combined search proposed here it is possible to substantially lower the threshold for a significant detection of feature models that contribute to both the power spectrum and the bispectrum.
In what follows it will be useful to introduce the notation
\begin{equation}
\bar{X}=\frac{\hat{X}}{\text{Var}\left[\hat{X}\right]^{\frac{1}{2}}}
\end{equation}
for any estimator or more generally random variable $\hat{X}$.

Let us start by quantifying the look-elsewhere effect in an individual survey. We will use a subscript $i=P,B$ to indicate that the discussion applies to both the power spectrum and the bispectrum. We are interested in the distribution of the maximum amplitude expected from fitting the noise, i.e. we study the distribution of
\begin{equation}
\bar{A}_i^{\text{max}}=\max\limits_{\phi_i,\omega}{\bar{A}_i(\omega,\phi_i)}
\end{equation}
under the null hypothesis that there are no underlying feature models. Of course we could use MC simulations to answer this question but rather than choosing this route we will try to derive an analytic model for this distribution that gives more insight and allows an easy generalisation to a joint significance later on.

To derive an analytic model we make two assumptions that should both be conservative:
\begin{itemize}
 \item[a)] The maximum amplitude at each $\omega$ given by $\max\limits_{\phi_i}\bar{A}_i(\omega, \phi_i)$ follows a chi distribution with two degrees of freedom for each value of $\omega$.
 \item[b)] Amplitude measurements at sufficiently separated frequencies are independent.
\end{itemize}
We can justify assumption (a) as follows. Neglecting minor effects coming from first finding the best-fit $\Lambda$CDM model in the power spectrum we can write the maximum amplitude at a given $\omega$ in both the power spectrum and the bispectrum as
\begin{equation}
\max\limits_{\phi_i}\bar{A}_i(\omega, \phi_i)=\max\limits_{\phi_i}\frac{\cos{\phi}\,\hat{X}_i+\sin{\phi}\,\hat{Y}_i}{\<\left(\cos{\phi}\,\hat{X}_i+\sin{\phi}\,\hat{Y}_i\right)^2\>^{\frac{1}{2}}}
\end{equation}
where $\hat{X}_i$, $\hat{Y}_i$ are Gaussian distributed. This simply follows from $\sin{(\omega k+\phi)}=\sin{(\omega k)}\cos{\phi}+\cos{(\omega k))}\sin{\phi}$ and the fact that the amplitude estimates are linear in the theoretical models. $\hat{X}_i$ corresponds to the estimate for the sine mode whereas $\hat{Y}_i$ corresponds to the estimate of the cosine mode. Now if we assume that $\<\hat{X}_i\,\hat{Y}_i\>=0$ we simply have
\begin{equation}
\max\limits_{\phi_i}\bar{A}_i(\omega, \phi_i)=\sqrt{\bar{X}_i^2+\bar{Y}_i^2}
\end{equation}
This is exactly a chi distribution with two degrees of freedom. Even though the assumption $\<\hat{X}_i\,\hat{Y}_i\>=0$ might not be exactly satisfied we expect correlations between the Gaussian random variables will change the chi distribution to something that should generally produce smaller $p$-values or equivalently assign larger significances to observed amplitudes. In this sense working with the chi distribution with two degrees of freedom is conservative. The second assumption makes a statement about the amount of independent information at different $\omega$. We know from Sec.~\ref{sec:Results} that correlations are generally not large for our stepsize in $\omega$ but obviously there are some. Especially if we decrease the step size in $\omega$ and use a finer grid the correlations between neighbouring bins will become more pronounced. We will later introduce a parameter for the effective number of independent bins that will account for these correlations and assume they are uncorrelated for now to derive an analytic expression.

\begin{figure}
 \centering
\includegraphics[scale=.23]{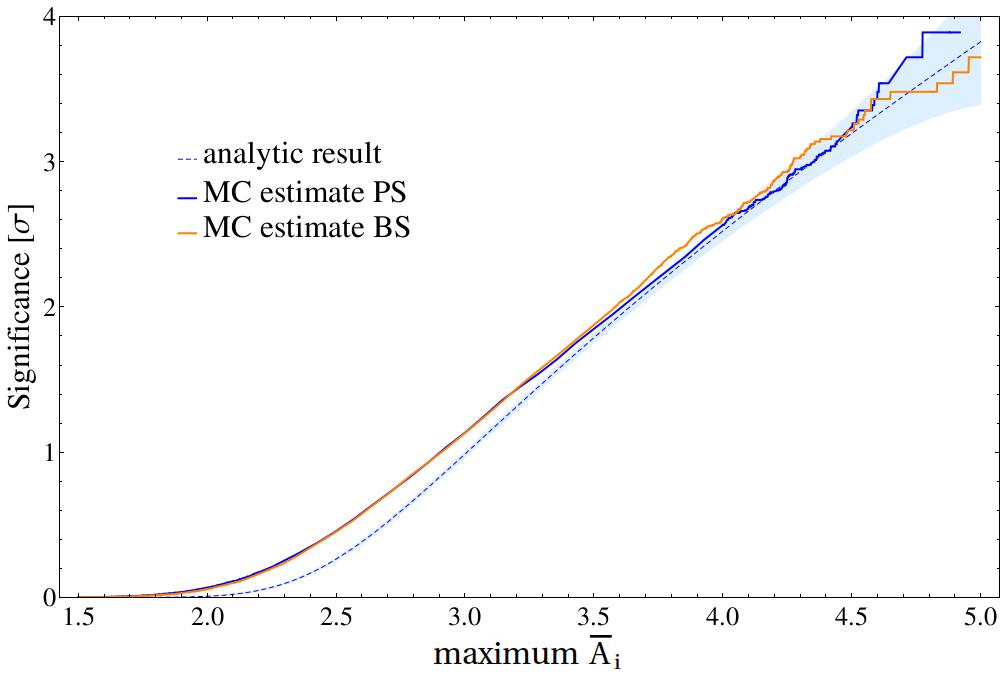}
\caption{Analytic prediction for the significance of results in the individual surveys compared to exact results from 10000 MC samples.}
\label{fig:sigdiagind}
\end{figure}

With these two assumptions we simply have
\begin{eqnarray}\nonumber
&\quad&P(\bar{A}_i^{\text{max}}\ge x)=1-P(\bar{A}_i^{\text{max}}<x)\\\nonumber
&=&1-\left(P(\bar{A}_i^{\phi}(\omega)<x)\right)^N=1-\left(F_{\chi,2}(x)\right)^N\\
&=&1-\left(F_{\chi^2,2}(x^2)\right)^N=1-\left(1-\exp{\left(-\frac{x^2}{2}\right)}\right)^N
\end{eqnarray}
where $N$ is the number of frequencies observed and $F_{\chi,2}$ and $F_{\chi^2,2}$ are the cumulative distribution functions of the chi distribution and chi-squared distribution with two degrees of freedom respectively. As mentioned above we can now introduce an effective number of independent frequencies $N_{\text{eff}}$ instead of the fixed number $N$ above to account for correlations between frequency bins. This parameter can then be chosen such that the analytic significances agree well with MC simulations. Rather than working with the $p$-values calculated above we translate them into significances $S$ in units of standard deviations of a normal distribution via
\begin{equation}\label{eq:ptosig}
S=2^{\frac{1}{2}}\text{Erf}^{-1}\left[1-p\right]\,.
\end{equation}
Figure \ref{fig:sigdiagind} compares the significances we obtained from 10000 MC runs with the significances predicted from our analytic model using $N_{\text{eff}}=35$.
This is a sensible value for $N_{\text{eff}}$ indicating that there are some dependencies among the 41 frequencies used in this survey.  $N_{\text{eff}}$ is related to the effective bin width $\Delta\omega_{\text{eff}}$ mentioned previously. The fact that $N_{\text{eff}}$ is less than the number of bins included agrees with $\Delta\omega_{\text{eff}}\gtrsim10$, a rough estimate that we based on the covariance matrices studied in Sec.~\ref{sec:Results}. The agreement between the analytic model and the MC results is good especially in the tail that we care about most when judging potentially interesting results. At lower significances the analytic model underestimates the exact result which is plausible given that our treatment does not account for correlations in an entirely rigorous fashion. 

For a combined survey the look-elsewhere effect depends on the number of free parameters of the model. In our case the most extreme look-elsewhere effect occurs if we allow both the relative amplitude $r$ and the relative phase $\phi_B-\phi_P$ to vary. The normalised combined amplitude is given by the optimal combined amplitude estimate introduced in the last section $\hat{A}$ divided by its variance which gives
\begin{equation}
\bar{A}=\frac{\hat{A}}{\text{Var}(\hat{A})}=\frac{\bar{A}_P+R\bar{A}_B}{(1+R^2)^\frac{1}{2}}
\end{equation}
where we introduced a variance weighted relative amplitude $R=r\,V_P^{\frac{1}{2}}/V_B^{\frac{1}{2}}$. Maximising with respect to $r$ is evidently the same as maximising with respect to $R$. We have to find the distribution of the maximum of this amplitude in a survey maximising with respect to $R$, $\phi_P$, $\phi_B$ and $\omega$
\begin{equation}
\bar{A}^{\text{max}}=\max\limits_{R, \phi_P, \phi_B, \omega}\bar{A}(\omega, \phi_P, \phi_B, R)
\end{equation}
Maximisation with respect to $R$ following exactly the same reasoning as before for $\phi$ in the individual surveys gives
\begin{equation}
\max\limits_{R}{\bar{A}}=\sqrt{\bar{A}_P^2+\bar{A}_B^2}
\end{equation}
Maximisation over the $\phi_i$ can be done individually giving
\begin{equation}
\max\limits_{R, \phi_P, \phi_B}\bar{A}=\sqrt{\bar{X}_P^2+\bar{Y}_P^2+\bar{X}_B^2+\bar{Y}_B^2}
\end{equation}
This is a chi distribution with four degrees of freedom and as before $p$-values for the full survey are given by
\begin{eqnarray}
\nonumber&\quad&P(\bar{A}^{\text{max}}\ge x)=1-\left(F_{\chi,4}(x)\right)^N=1-\left(F_{\chi^2,4}(x^2)\right)^N\\
&=&1-\left(1-\exp{\left(-\frac{x^2}{2}\right)}-\frac{x^2}{2}\exp{\left(-\frac{x^2}{2}\right)}\right)^N
\end{eqnarray}
Actual theories might predict a specific phase relation so that there is less opportunity for a look-elsewhere effect. To study this possibility let us simply assume $\phi_P=\phi_B=\phi$ and look for the maximum amplitude found for any common phase $\phi$ so that in this case
\begin{equation}
\bar{A}^{\text{max}}=\max\limits_{R, \phi, \omega}\bar{A}(\omega, \phi, R)
\end{equation}
Obviously we are likely to be faced with more complicated phase relations in searches for feature models but this should have little effect on the look-elsewhere effect. We found that in this case the resulting distribution is not a simple chi distribution with three degrees of freedom as one might expect based on the previous cases. However, we show in App.~\ref{app:distphaserel} that we can approximate the resulting cumulative distribution function (CDF) $F(x)$ of the exact distribution well with
\begin{equation}
F(x)\sim F_{\chi,3.5}(x)
\end{equation}
where we assume for simplicity $\<\hat{Y}_P^2\>/\<\hat{X}_P^2\>\approx\<\hat{Y}_B^2\>/\<\hat{X}_B^2\>$.

\begin{figure}
 \centering
\includegraphics[scale=.23]{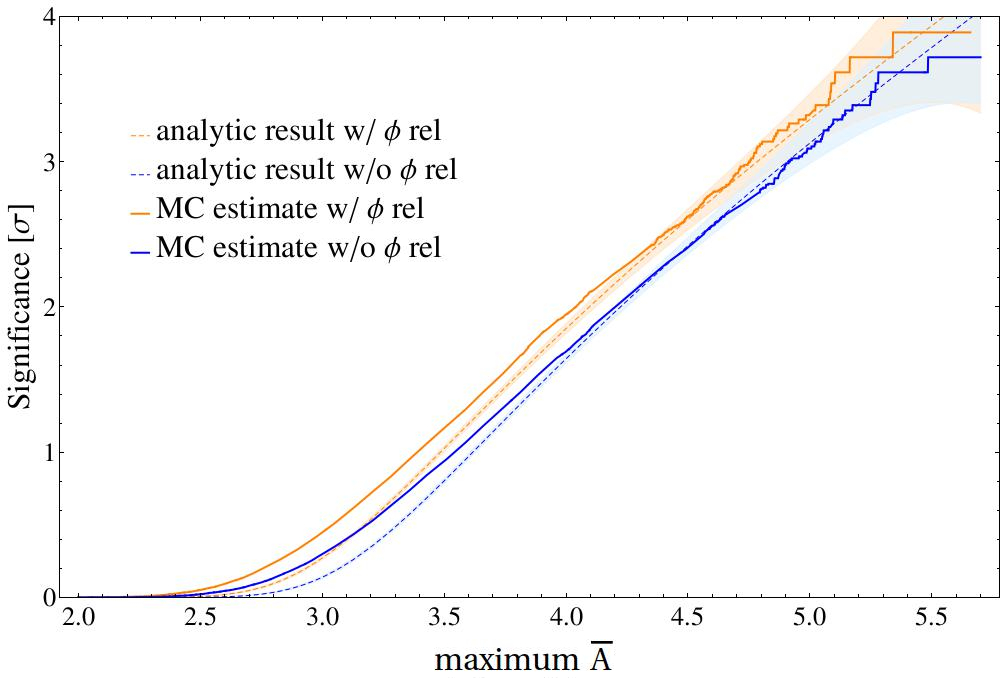}
\caption{Analytic predictions for the significance of results in the combined survey compared to exact results from 10000 MC samples. We show the case of not assuming a phase relationship between the power spectrum and bispectrum and also the case of assuming $\phi_P=\phi_B=\phi$.}
\label{fig:sigdiagcomb}
\end{figure}

Figure \ref{fig:sigdiagcomb} compares the significances we obtained from 10000 MC runs with the significances predicted from our analytic models in both cases using $N_{\text{eff}}=35$. Again, the agreement is good, particularly in the most interesting region of high significance.
It is evident that the look-elsewhere effect resulting from not assuming a phase relation is rather small and the joint signficances assigned to findings are very similar in both cases.

\begin{figure}
 \centering
\includegraphics[scale=.4]{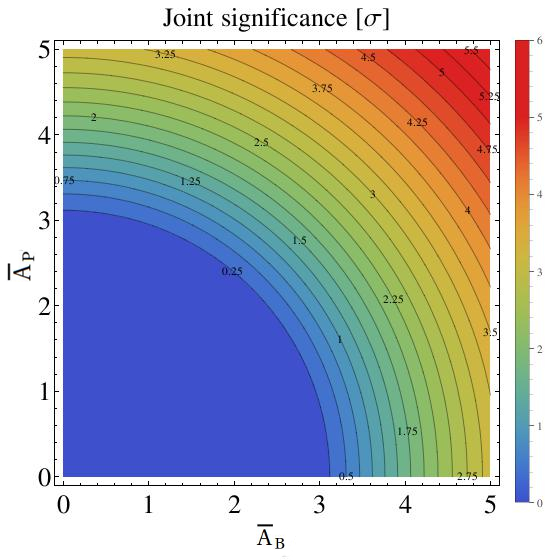}
\caption{Contour plot of the look-elsewhere adjusted significances of finding amplitudes $\bar{A}_P$ and $\bar{A}_B$ at the same $\omega$.}
\label{fig:sigcont}
\end{figure}

Summing up the results of this section, we conclude that a combined search for a single frequency model can reveal interesting results even if the power spectrum and bispectrum show no statistically significant signals on their own. If we find a pair of amplitudes $\bar{A}_P$ and $\bar{A}_B$ at some frequency $\omega$ we can assign it a significance based on the expectation for fitting a featureless Gaussian realisation. The look-elsewhere adjusted joint significance $S$ (in units of sigma) in the most conservative case, i.e. assuming no phase relation and maximising over the relative amplitude, is then given by
\begin{eqnarray}
S&=&2^{\frac{1}{2}}\text{Erf}^{-1}\left[1-P\left(\bar{A}^{\text{max}}>(\bar{A}_P^2+\bar{A}_B^2)^{1/2}\right)\right]\\
&=&2^{\frac{1}{2}}\text{Erf}^{-1}\left[\left(F_{\chi^2,4}\left(\bar{A}_P^2+\bar{A}_B^2\right)\right)^{N_{\text{eff}}}\right]
\end{eqnarray}
The resulting contour plot using $N_{\text{eff}}=35$ is shown in Fig.~\ref{fig:sigcont}. 
As an example if we find matching 3.5 sigma peaks at some frequency $\omega$ in both the power spectrum and the bispectrum then their joint significance from Fig.~\ref{fig:sigcont} is at the 3.1 sigma level while the two results on their own are at the 1.8 sigma level as is evident from Fig.~\ref{fig:sigdiagind}. Note that the possible boost in significance can increase with the range of the survey as the chances of the largest peaks in the noise occurring at the same $\omega$ decreases. An advantage of having analytic expressions for the look-elsewhere adjusted significances is that we can easily extrapolate to obtain predictions for surveys covering a much larger $\omega$ range. For an $\omega$ range of $\sim 400$ we obtained $N_{\text{eff}}\sim 35$. Recent surveys of oscillations in the power spectrum (for example \cite{Meerburg:2013SearchOscP1, Meerburg:2013SearchOscP2}) cover frequencies up to $\omega\sim\mathcal{O}(10^4)$ which should be equivalent to $N_{\text{eff}}\sim \mathcal{O}(10^3)$. A 1.8 sigma result in an individual survey with $N_{\text{eff}}=1000$ requires a 4.35 sigma peak at some $\omega$. Matching peaks in both surveys at this level now give a 3.8 sigma result. This shows how bigger boosts in significance are possible for higher $N_{\text{eff}}$.

\begin{figure}
 \centering
\includegraphics[scale=.4]{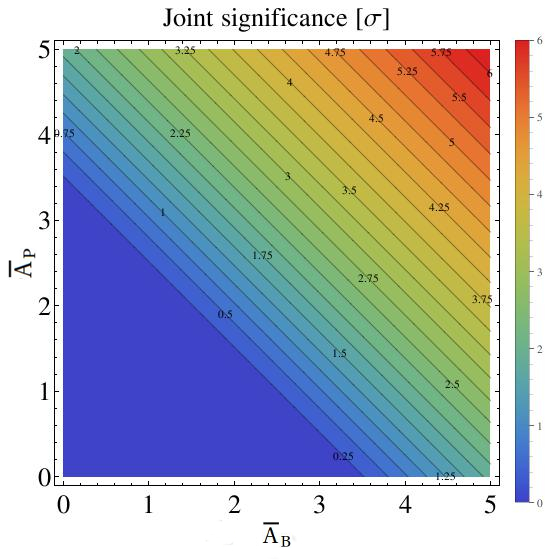}
\caption{Contour plot of the look-elsewhere adjusted significances of finding amplitudes $\bar{A}_P$ and $\bar{A}_B$ at the same $\omega$ in a survey where we only consider models with $R=1$ and $\phi_P=\phi_B$.}
\label{fig:sigcontfixed}
\end{figure}

It is of some interest to compare the obtained significances to a scenario where we are simply given twice as much data and combine two power spectrum or bispectrum surveys. Given the approximate independence of the power spectrum and bispectrum amplitude estimates discussed in the previous sections, this situation is clearly equivalent to a combined search where we are only looking for models that predict variance weighted amplitude ratios $R=1$ and a fixed phase relation $\phi_P=\phi_B$. It also applies to some extent to combining temperature and polarisation measurements where we generally expect very robust theoretical predictions relating features in the temperature and polarisation power spectra. Obviously this is a simplified statement as it neglects that temperature and polarisation fluctuations are correlated and also assumes that the S/N of the feature models is comparable which is not necessarily the case. We will briefly discuss the inclusion of polarisation in Sec.~\ref{subsec:Polarisation}. For $R=1$ and $\phi_P=\phi_B$ the contour plot for the resulting significances is shown in Fig.~\ref{fig:sigcontfixed}.
As we significantly constrained the range of models under consideration there is less opportunity for noise to produce apparently significant results meaning that finding large amplitudes in both surveys gives a more significant look-elsewhere adjusted result. However, finding a large signal in one survey and much less signal in the other gives a less significant result as this would be evidence against a model with $R=1$.

While there can be some loss of significance involved in considering models with different $R$ and relative phases it seems hard to motivate a strong theoretical prior for these quantities. If anything, we expect $R<<1$ for most models. Thus it is more reasonable to consider the general class of models allowing $R$ to vary and we saw that gains in significance can still be made. This is encouraging but we can clearly not expect highly significant results after a proper look-elsewhere correction if the individual surveys show no interesting look-elsewhere adjusted results at all. In the example above we assumed that there are 1.8 sigma results above the expectation from noise present in both surveys which is already a relatively big signal. However, there is the possibility that significant evidence for feature models can be found pursuing various other routes. We will briefly discuss different options in the next section.

\section{Refining the search for oscillatory features}
\label{sec:RefinedSearch}
\subsection{Multifrequency models}
\label{subsec:MultiStat}
Some feature models can generate modulations at multiple, well-separated, frequencies. This could be due to multiple sharp features at different locations in the potential or some other mechanism (see for example \cite{Chen:PrimFeatures, Chen:StandardClock}). Looking for these types of models is sensible both in the context of individual surveys as well as combined searches. We will introduce a useful statistic for multifrequency models focusing on a combined search but the results readily apply to the individual surveys with appropriate changes as discussed below. As a phenomenological model for this class of features we take
\begin{eqnarray}
P(k) &\sim& 1+\sum\limits_{i=1}^{M}A_{P,i}\sin(\w_i (2k) + \o_{P,i})\\
B(k_1,k_2,k_3) &\sim&\sum\limits_{i=1}^{M}A_{B,i}\sin(\w_i K + \o_{B,i})
\end{eqnarray}
where as usual $K=k_1+k_2+k_3$ and the sum is over the $M$ different frequencies contributing to the model. This model is clearly very general. At each frequency $\omega$ we allow for any combination of phases and any relative amplitude. If the features are for example due to well-separated steps in the potential we do not expect their properties to be related so that this is a reasonable assumption.

Let us assume for a moment that the survey is made up of $N_{\text{eff}}$ uncorrelated frequency bins rather than an arbitrary number of correlated bins that depends on the resolution of the $\omega$ grid used in the survey. Following the same reasoning as in the previous sections for a given $M$, the maximum significance for any optimal estimate for the overall amplitude is then
\begin{equation}
\bar{A}^{\text{max}}_M=\left(\sum\limits_{i=1}^{M}(\bar{A}_{P,i}^2+\bar{A}_{B,i}^2)\right)^{\frac{1}{2}}
\end{equation}
where as before $\bar{A}_{P,i}$ and $\bar{A}_{B,i}$ are the maximum normalised amplitudes found in the power spectrum and bispectrum at a given $\omega_i$ and the set of frequencies $\omega_i$ is obtained by picking the $M$ largest $\bar{A}_{P}^2+\bar{A}_{B}^2$ amongst all frequencies. A sensible way to assign a significance to an amplitude $\bar{A}^{\text{max}}_M$ in this case is a two-step process and rather tedious. First we obtain the distributions of amplitudes $\bar{A}^{\text{max}}_M$ for each choice of $M$. Comparing a given value of $\bar{A}_M$ to this distribution produces a significance $\sigma_M$. This significance indicates how unlikely it is to find an optimal amplitude estimate this large given a model with $M$ distinct frequencies in a featureless realisation of the data. However, $M$ is a parameter that we can choose freely so we need to take another look-elsewhere effect into account. We do so by comparing the $\sigma_M$ to the distribution of
\begin{equation}
\sigma^{\text{max}}=\max\limits_{M}\sigma_M
\end{equation}
This procedure assigns a total look-elsewhere adjusted significance to a given multifrequency model.

While this method is a rigorous and reliable measure of significance for multiple peaks in the data, it is rather tedious in practice due to the need of having to calculate the distributions for every possible $M$ which can only be done accurately using a vast amount of MC simulations. It is also complicated to incorporate correlations between frequency bins in an actual survey. These correlations have to be accounted for in the construction of the optimal estimator for the amplitude of a given model. We thus investigated whether there is a simpler way to arrive at significances that approximately reproduce the results of the procedure outlined above for the case of uncorrelated bins and allow for an easy generalisation.

A simple and rather natural measure that is sensitive to the existence of multiple large amplitudes in a survey is obtained in the following way. We simply take the root of the sum of squares of the significances assigned to each frequency $\omega$ using the method described in Sec.~\ref{subsec:CombStat} for the case of maximising over relative amplitudes and phases. More precisely we define the integrated statistic $S_I$ given by
\begin{equation}
S_I^2=\frac{\Delta\omega}{\Delta_{\text{eff}}}\sum\limits_{\omega}2\,\text{Erf}^{-1}\left[\left(F_{\chi^2,4}\left(\bar{A}_{P,\omega}^2+\bar{A}_{B,\omega}^2\right)\right)^{N_{\text{eff}}}\right]^2
\end{equation}
and obtain significances for a given survey by comparison to the distribution of this statistic. Here, $\Delta\omega$ is the stepwidth of the frequency grid and $\Delta_{\text{eff}}$ is the effective stepwidth that we define as 
\begin{equation}
\Delta_{\text{eff}}=\frac{\omega_{\text{max}}-\omega_{\text{min}}}{N_{\text{eff}}-1}
\end{equation}
In our case we have $\Delta_{\text{eff}}\approx 11.5$. Note that in the expression for $S_I^2$ $\bar{A}_{P,\omega}$ and $\bar{A}_{B,\omega}$ are the maximum normalised amplitudes found at a given $\omega$. While it might not be clear at first sight that the integrated statistic is a roughly equivalent way of assessing evidence for multifrequency models in a survey, we show in App.~\ref{app:MultiStat} that this is indeed the case for uncorrelated bins as long as we are mainly looking for models with a relatively small number of frequencies $M$.

\begin{figure}
 \centering
\includegraphics[scale=.23]{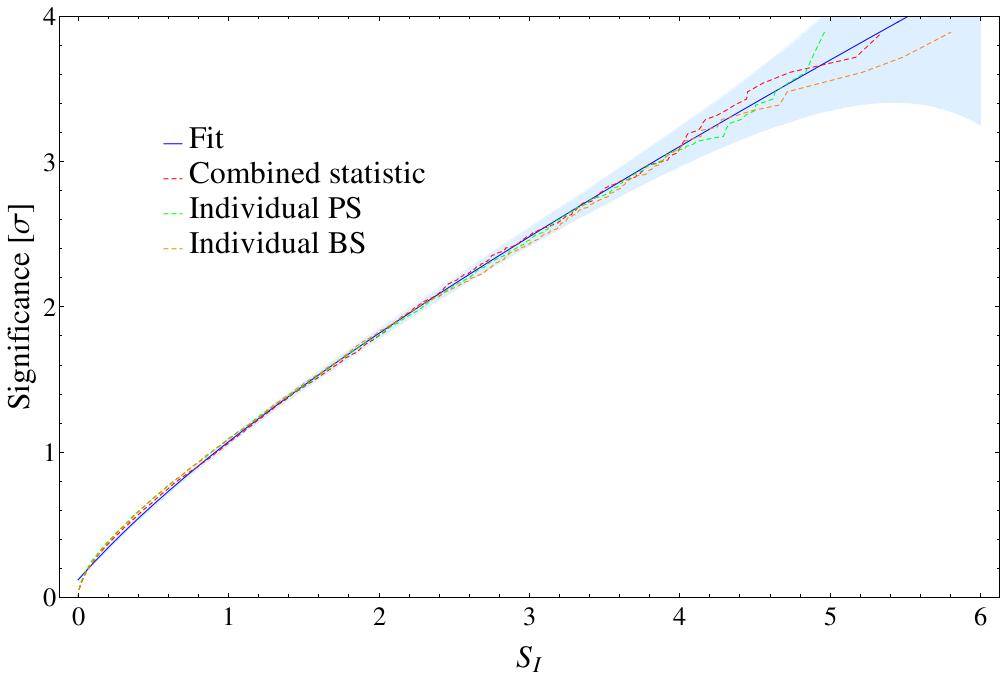}
\caption{Distribution of the integrated statistics $S_I$ for the individual and combined surveys given as significances. As discussed in the main text the distributions are nearly identical and well fit with a simple expression.}
\label{fig:intstats}
\end{figure}

The same reasoning applies to the individual surveys giving rise to a corresponding multipeak statistic
\begin{equation}
S_I^2=\frac{\Delta\omega}{\Delta_{\text{eff}}}\sum\limits_{\omega}2\,\text{Erf}^{-1}\left[\left(F_{\chi^2,2}\left(\bar{A}_{i,\omega}^2\right)\right)^{N_{\text{eff}}}\right]^2
\end{equation}
where as before $i=P,B$. These multifrequency statistics have a number of interesting properties. The distributions of $S_I$ for the individual and combined surveys are shown in Fig.~\ref{fig:intstats} where we plotted significances obtained via Eq.~(\ref{eq:ptosig}) rather than the distributions themselves.
We see that the distributions are almost identical. This is somewhat unsurprising given that we sum over significances. The CDFs for the statistics $F_{S_I}$ are all well fit using the simple expression
\begin{equation}
F_{S_I}(x)=1-\exp\left(-(a\,x^2+b\,x+c)\right)
\end{equation}
with $a=.0921$, $b=.8762$ and $c=.1022$ for $x>>0$ so that we can confidently extrapolate to study highly significant results in the tail of the distributions.

On top of being a good measure of whether or not there is evidence for multifrequency models in the data as discussed above, we find that the the scaling with the survey range is largely absorbed in the parameter $N_{\text{eff}}$. This is convenient as results for different survey ranges can be simply compared to the distribution in Fig.~\ref{fig:intstats} as a first check rather than having to obtain a distribution from a large number of MC simulations for each case. Furthermore the statistic has a well-defined infinite resolution limit. Taking $\Delta \omega\rightarrow 0$ the statistic $S_I$ simply becomes an integral where $N_{\text{eff}}$ takes its asymptotic value. We evaluated the statistics in this limit and found that $N_{\text{eff}}$ increases, or equivalently $\omega_{\text{eff}}$ decreases, by a small amount $<10\%$ compared to the value obtained with stepwidth $\Delta\omega=10$ that we used above. Hence, a stepwidth of $\Delta\omega\sim10$ is already close to the infinite resolution limit. Moreover the distributions of the statistics $S_I$ are virtually identical to those plotted in Fig.~\ref{fig:intstats}. 

\begin{figure}
 \centering
\includegraphics[scale=.23]{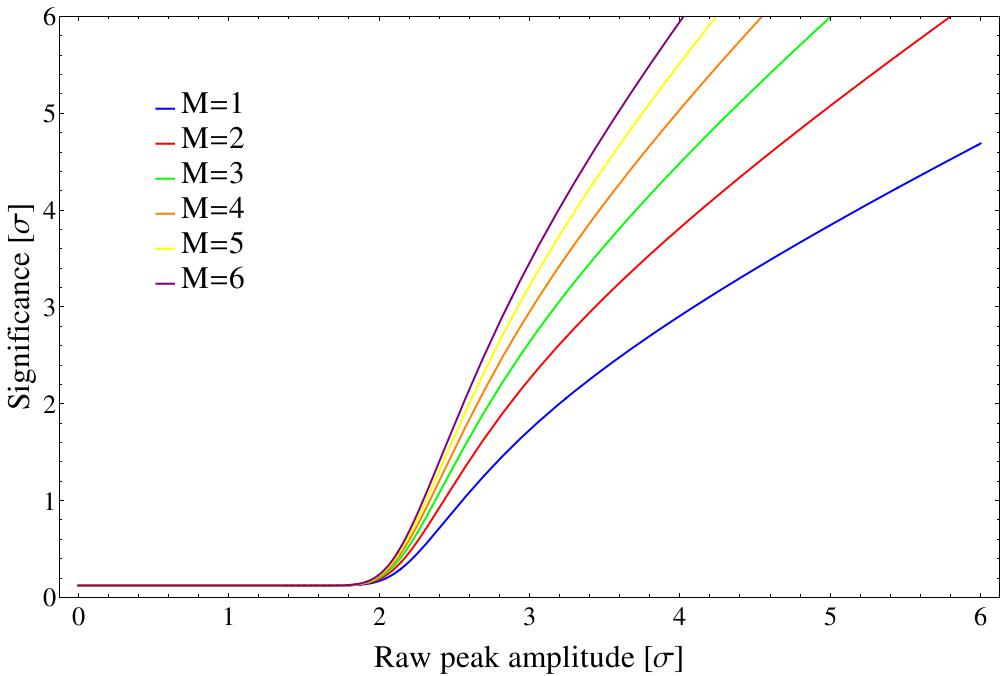}
\caption{Estimate of the significances assigned to pairs of peaks of a given amplitude at $M$ frequencies $\omega$ in a combined survey for various values of $M$.}
\label{fig:multipeaksigs}
\end{figure}

Having constructed a sensible statistic to search for multiple peaks it is useful to estimate how many peaks at a given height are needed to present statistically significant evidence for a multifrequency model. For this purpose we simply assume that there are $M$ peaks of a given amplitude $\sigma$ in both the power spectrum and bispectrum, each with width $\Delta\omega_{\text{eff}}$. Simply setting contributions from all other frequencies to zero we then calculate the significance assigned to this realisation using the integrated statistic $S_I$. The results are shown in Fig.~ \ref{fig:multipeaksigs}.
While the appearance of multiple peaks with low amplitudes does not greatly enhance the overall significance, several marginally significant results at different $\omega$ can constitute strong evidence for multifrequency models. 

\subsection{Incorporating higher order correlators}
Another possibility to increase the S/N for a given feature model is to include observations of higher order correlators. Of course, this strategy is limited to models that produce not only a large observable bispectrum but also observable higher order correlation functions. As an example it is conceivable to include measurements of the amplitude of the contribution to the connected part of the four-point correlator, the trispectrum of the feature model $T$, using the estimator \cite{Regan:Trispec}
\begin{widetext}
\begin{eqnarray}\nonumber
\hat{A}_T&=&\frac{1}{N_T}\sum_{\ell^{\phantom{'}}_i \ell'_i m^{\phantom{'}}_i m'_i}  T^{l_1l_2l_3l_4}_{m_1m_2m_3m_4} (C^{-1})_{\ell^{\phantom{'}}_1\ell'_1m^{\phantom{'}}_1m'_1} (C^{-1})_{\ell^{\phantom{'}}_2\ell'_2m^{\phantom{'}}_2m'_2} (C^{-1})_{\ell^{\phantom{'}}_3 \ell'_3 m^{\phantom{'}}_3 m'_3}(C^{-1})_{\ell^{\phantom{'}}_4 \ell'_4 m^{\phantom{'}}_4 m'_4}\\
&\quad&\qquad\qquad\qquad\qquad\left(a_{\ell'_1 m'_1} a_{\ell'_2 m'_2} a_{\ell'_3 m'_3} a_{\ell'_4 m'_4}-6\,\<a_{\ell'_1 m'_1} a_{\ell'_2 m'_2}\> a_{\ell'_3 m'_3} a_{\ell'_4 m'_4}+ 3\<a_{\ell'_1 m'_1} a_{\ell'_2 m'_2}\> \<a_{\ell'_3 m'_3}a_{\ell'_4 m'_4}\>\right)\,,
\end{eqnarray}
\end{widetext}
where $N_T$ is the appropriate normalisation factor rendering the estimator unbiased. Note that this estimator is uncorrelated with both the power spectrum and bispectrum amplitude estimators as can be checked easily. Based on the intuition gained from studying the case of combining amplitude measurements from the former two, we expect this estimator to produce statistically nearly independent constraints for large enough $l_{\text{max}}$. In the same sense as discussed above we could interpret this as a consequence of the multivariate CLT stating that uncorrelatedness implies asymptotic independence in the large sample limit. The optimal estimator for the overall amplitude of a given feature model is then a linear combination of these three amplitude estimators with coefficients determined by the relative amplitudes and the variances of the estimators just as we discussed in Sec.~\ref{sec:CombSearch}.

\subsection{Including Polarisation}
\label{subsec:Polarisation}
Oscillatory primordial features are clearly not only imprinted in the temperature fluctuations (T) but should also be visible in the polarisation of the CMB, in particular the E-mode fluctuations (E). Both the fluctuations in temperature and polarisation originate from the same primordial fluctuations and are simply convolved with different transfer functions to obtain the observed CMB anisotropies. Thus any model that exhibits a TT power spectrum modulation necessarily also produces a modulation in the EE power spectrum and the TE cross-spectrum. In this sense polarisation is a more stringent test for oscillatory features and there is no need to include additional tunable parameters like a relative amplitude $R$ discussed above in a combined temperature and polarisation search.

Another very attractive feature of polarisation measurements is that the polarisation transfer functions are narrower due to projection effects \cite{Hu:PrincipalPower}. In the case of temperature a fluctuation with a given wave vector $k$ nearly parallel to the line of sight will not only contribute to fluctuations with $l\sim k\Delta\eta$ but also contribute to $l<<k\Delta\eta$. This effect is much less prominent for polarisation and the projection is sharper. For features that oscillate with high frequency this means that we generally expect the damping of the oscillation due to the convolution with the transfer functions to be less severe and signals should stand out more in the polarisation data (see for example \cite{Mortonson:PolFeatures}).  

\begin{figure}
 \centering
\includegraphics[scale=.23]{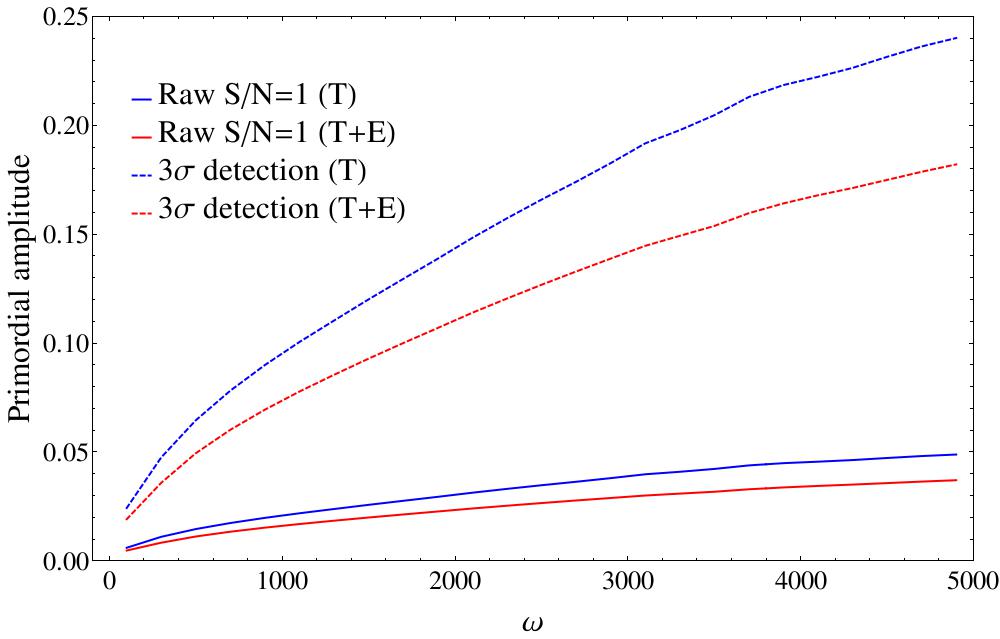}
\caption{A rough estimate of the primordial oscillation amplitude needed to obtain $S/N=1$ (solid) and a 3-sigma look-elsewhere adjusted detection (dashed) in a Planck-like experiment using only power spectrum constraints. Both T only and T+E surveys are considered where we crudely estimated the foreground and noise contamination of E mode to be twice that of the temperature data.}
\label{fig:SNforecast}
\end{figure}

However, polarisation measurements are complicated by foregrounds and typically only a smaller $l$ range is accessible. Furthermore the information in the polarisation data is not entirely independent. There are strong correlations between the E and T fluctuations that reduce the amount by which we can hope to improve the S/N of feature models. Figure \ref{fig:SNforecast} shows a rough estimate what we can expect from a joint T and E analysis.
The figure assumes 40\% sky coverage%
\footnote{We account for incomplete sky coverage simply by multiplying with $f_{\text{sky}}=.4$. This neglects the fact that masking introduces correlations between nearby multipoles. These correlations can further decrease the signal to noise when looking for oscillatory modulations of the power spectrum.}%
and uses noise levels typical for the Planck mission while simply taking $N^E_l\sim 2 N^T_l$ (for the purpose of this simple calculation foregrounds can be thought of as being included in the noise power spectra). Cross-correlations are accounted for but have a relatively small effect. Several points should be highlighted. First of all this figure clearly shows how it is generally extremely difficult to observe high-frequency oscillations in the CMB. Assuming that the oscillations extend over the entire $l$-range we would naively expect an amplitude $A\sim l_{\text{max}}^{-1}=\mathcal{O}(10^{-3})$ to produce S/N of unity. However, the smoothing due to the transfer functions as well as lensing of the power spectra suppresses the amplitude by orders of magnitude. As is indicated by the dashed lines in the figure, we would need primordial oscillations with amplitudes $\mathcal{O}(.1)$ and higher for a significant detection at high $\omega$ after the look-elsewhere effect is taken into account. Also, if the feature originated for example from steps in the potential that are not infinitely sharp we have to take into account that modes that were deep in the horizon at the time of the feature are unaffected. Thus the oscillation is damped for larger $k$ and does not extend over the full $l$-range. This would lead to further considerable reduction of the S/N of the feature.

Even though we can only observe polarisation well over a much smaller $l$-range due to higher noise levels and foregrounds as well as a much smaller signal, the fact that the polarisation spectra are less affected by smoothing makes significant gains in S/N for feature models possible. Obviously this conclusion depends on how badly foregrounds affect the polarisation signal.

\section{Summary and Conclusions}
Searches for oscillatory feature models in the CMB have been undertaken focusing on both the power spectrum and the bispectrum. In both cases various interesting oscillation scales have been identified. Given that feature models typically predict a power spectrum modulation as well as a bispectrum the question arises of how to combine results. This is particularly interesting given the fact that we generally expect the oscillation scales predicted by theory to be closely related so that finding evidence of an oscillation at the same scale in both the power spectrum and the bispectrum could present us with evidence for a feature model. However, before we can get excited about matching signals at a given $\omega$ we need to investigate possible dependencies between measurement based on the power spectrum and bispectrum. Given that the measurements are based on the same set of data at some level these dependencies are expected to exist.

Evaluating the quadratic power spectrum estimator for the feature model amplitude and the corresponding bispectrum estimator on the same set of CMB simulations we showed that there is no evidence of correlation between the absolute values of the amplitudes. In particular we found no evidence for any correlations using $10^4$ simple noiseless full-sky simulations which implies that correlations must be below the \% level in this case. We also studied the case of incomplete sky coverage and anisotropic coloured noise, evaluating the estimators on 400 more realistic CMB realisations. Again no correlations between the amplitudes above the noise level were observed, implying that masking and anisotropic noise do not introduce significant correlations. To exclude the possibility of more complicated dependencies we also calculated the distance correlation for our sets of simulations. This is a measure of statistical dependence that only vanishes for statistically independent quantities. We observed no evidence of dependencies at the level we could resolve with the amount of simulations available to us. To support this conclusion we provided an analytic estimate for the simplest scenario suggesting that any dependencies between the amplitudes should be suppressed by a factor of $l_{\text{max}}^{-1}$. For the case of Planck resolution with $l_{\text{max}}\sim 2000$ that we studied, this suggests that any correlations and other dependencies should be of the order $10^{-3}$. This is clearly consistent with the results of the simulations as correlations of this order are far below the MC noise level expected from $10^4$ samples. We conclude that any dependencies between measurements of feature models in the power spectrum and bispectrum are very nearly statistically independent. Any dependencies can be ignored to a very good approximation in a combined search for feature models. This is a key result of this paper.

Building on this conclusion we proposed an optimal amplitude estimator for a combined search using both the power spectrum and the bispectrum. Given that it is generally hard to construct an accurate likelihood that incorporates higher order correlators we believe that this is the most natural statistic to consider when attempting to combine power spectrum and bispectrum estimators.

We went on to study the look-elsewhere effect in feature model surveys. For the generic model that we focused on in this work the distribution of amplitude estimates can be largely modelled analytically. The results can be applied to the individual surveys as well as a combined search. While we refer the reader to the main text for details we highlight that there are generally large look-elsewhere effects involved when searching for feature models. As we vary the frequency parameter $\omega$ amplitude estimates for different feature models become largely independent as soon as they are separated by more then an effective frequency step-size $\Delta\omega_{\text{eff}}$ which is mainly set by the resolution limit of the CMB experiment $l_{\text{max}}$. For generic models considered here at Planck resolution we have $\Delta\omega_{\text{eff}}\sim 10$. This implies that a typical survey covering a frequency range of order $10^3-10^4$ tests a large $\mathcal{O}(10^3)$ number of independent models. This implies that we expect to see naively very significant results $>4\sigma$ in a majority of realisations of a featureless Gaussian CMB even for the simplest class of models. This has to be taken into account properly to judge the significance of results. While we focused on look-elsewhere effects coming from frequency, phase and relative amplitude parameters, more complicated models can also have tuneable envelope parameters for example. This will exacerbate this problem.

Generalising the model to allow for multiple frequencies $\omega$ we constructed a simple integrated statistic $S_I$ that performs well at picking up on evidence for these kinds of models. The statistic gives rise to adjusted significances that allow for a reasonable judgement of whether or not the observation of large amplitudes at multiple frequencies in the data should be considered interesting.

The approximate independence of the feature model estimates based on the power spectrum and bispectrum should also extend to higher order correlators. For models that predict observable higher correlation functions this provides opportunity for further, more stringent, tests following the spirit of this work.

The Planck polarisation data is being released in due course and will provide us with yet another powerful tool to constrain feature models.

\begin{center}
\textbf{Acknowledgements}
\end{center}
We are very grateful to Benjamin Wallisch for many discussions and input, and with whom this work is being pursued further in an forthcoming publication on WMAP and Planck data. We also thank Xingang Chen, Yi Wang and Daniel Baumann for valuable discussions and comments. We are grateful to Juha J\"{a}ykk\"{a} and James Briggs for outstanding computational support. H.G. gratefully acknowledges the support of the Studienstiftung des deutschen Volkes and an STFC studentship. This work was supported by an STFC consolidated Grant No. ST/L000636/1. It was undertaken on the COSMOS Shared Memory system at DAMTP, University of Cambridge operated on behalf of the STFC DiRAC HPC Facility. This equipment is funded by BIS National E-infrastructure capital Grant No. ST/J005673/1 and STFC Grants No. ST/H008586/1 and No. ST/K00333X/1. E.P.S. thanks the MIAPP for their hospitality. We acknowledge use of the HEALPix package \cite{Gorski:HEALPix}.

\appendix
\section{Distribution of $\bar{A}^{\text{max}}$ for a fixed phase relation}
\label{app:distphaserel}
We assume a fixed phase relation $\phi_P=\phi_B=\phi$ and maximise with respect to the common phase $\phi$ so that
\begin{equation}
\bar{A}^{\text{max}}=\max\limits_{R, \phi, \omega}\bar{A}(\omega, \phi, R)
\end{equation}
As in the other cases we assume that the amplitudes for the sine and cosine modulations are independent at each omega. Furthermore we assume for simplicity $\<\hat{Y}_P^2\>/\<\hat{X}_P^2\>\approx\<\hat{Y}_B^2\>/\<\hat{X}_B^2\>$ and obtain
\begin{widetext}
\begin{equation}
\max\limits_{R, \phi}\bar{A}=\left(\frac{1}{2}\left(\bar{X}_P^2+\bar{Y}_P^2+\bar{X}_B^2+\bar{Y}_B^2+\sqrt{\left((\bar{Y}_P+\bar{X}_B)^2+(\bar{X}_P-\bar{Y}_B)^2\right)\left((\bar{Y}_P-\bar{X}_B)^2+(\bar{X}_P+\bar{Y}_B)^2\right)}\right)\right)^{\frac{1}{2}}
\end{equation}
\end{widetext}
The $\bar{X}_i$ and $\bar{Y}_i$ are independent Gaussian random variables with unit variance. This is not simply a chi distribution with three degrees of freedom as one might expect based on the cases discussed in the main text. However, we find that we can fit the probability density function (PDF) of this distribution, $f(x)$, well with the PDF of the chi distribution with 3.5 degrees of freedom, $f_{\chi,3.5}$.
\begin{figure}
 \centering
\includegraphics[scale=.23]{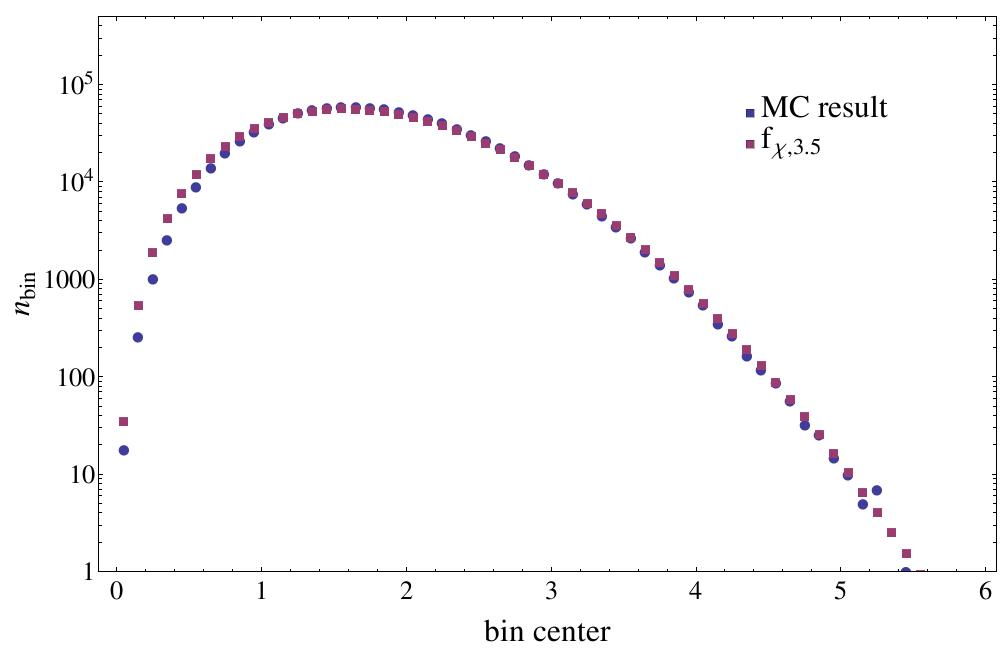}
\caption{Exact distribution obtained from $10^6$ random samples compared to a prediction based on the PDF of a chi distribution with 3.5 degrees of freedom.}
\label{fig:distfit_phaserel}
\end{figure} 
Figure \ref{fig:distfit_phaserel} compares this analytic approximation with the exact distribution obtained from $10^6$ random samples. This implies that we can simply approximate the CDF of the exact distribution as
\begin{equation}
F(x)\sim F_{\chi,3.5}(x)
\end{equation}
to extract the significances.

\section{Multi-frequency statistics}
\label{app:MultiStat}
We outlined a rigorous procedure for assigning significances to amplitudes of multifrequency models found in the data in Sec.~\ref{subsec:MultiStat}. The purpose of this appendix is to show that the much simpler integrated statistic $S_I$ from Sec.~\ref{subsec:MultiStat} reproduces the significances to good accuracy for moderate values of the number of frequencies of the model $M$. To show this we assume that there are $M$ peaks with height $\sigma$ at the same frequencies in both the power spectrum and the bispectrum in the data. A corresponding $M$-frequency model will give an amplitude estimate with raw significance $(2\,M\,\sigma^2)^{\frac{1}{2}}$ that needs to be compared to the distribution of maximum amplitudes for an $M$-frequency model just from noise to obtain a significance $\sigma_M$. As discussed in the main text this needs to be adjusted again to account for the fact that $M$ is a free parameter of the model to obtain a final significance $\sigma$. The corresponding analysis using the integrated statistic $S_I$ simply involves calculating the $p$-value for the value of the statistic obtained for $M$ peaks with height $\sigma$ given by
\begin{equation}
S_I=2\,M\,\text{Erf}^{-1}\left[\left(F_{\chi^2,4}\left(2 \sigma^2\right)\right)^{N_{\text{eff}}}\right]^2
\end{equation}
Note that this way of calculating the value of $S_I$ neglects contributions from other frequencies that occur in every survey. These contributions are typically very small and do not have a significant effect on the significance assigned to a set of large peaks. We performed both ways of assigning significances based on $10^4$ realisations of a mock survey that we obtained by drawing $N_{\text{eff}}=35$ independent samples from a chi distribution of degree 4 corresponding to the values of $(\bar{A}_P^2+\bar{A}_B^2)^{\frac{1}{2}}$ for each of the $N_{\text{eff}}=35$ independent $\omega$ bins in the survey.  
\begin{figure}
\centering
\includegraphics[scale=.23]{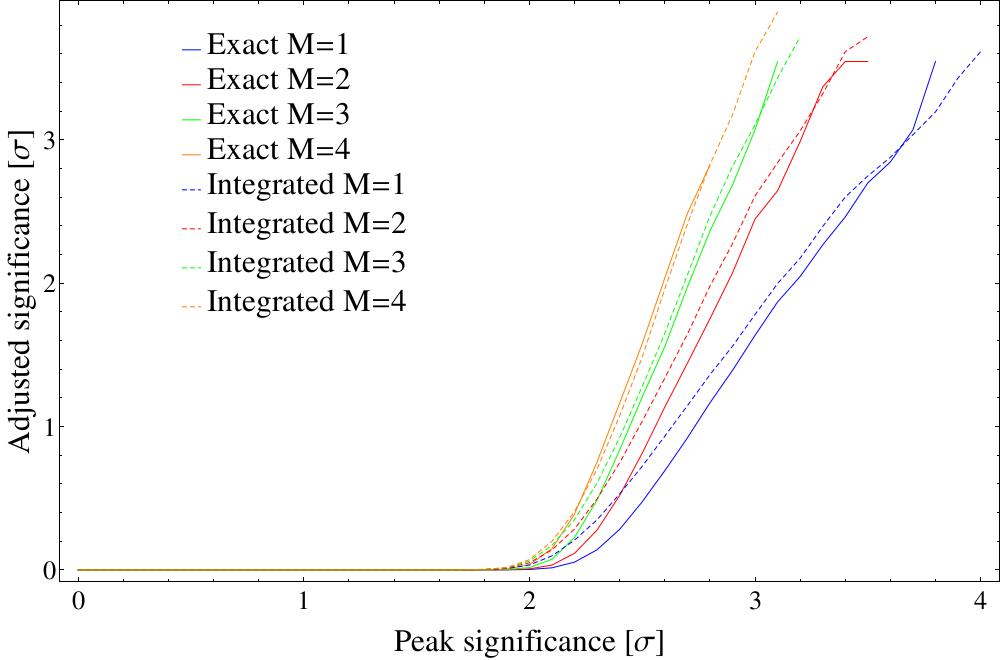}
\caption{Significances assigned to $M$ peaks of height $\sigma$ at the same frequencies in both the power spectrum and the bispectrum using a rigorous look-elsewhere analysis and the simple integrated statistic. The significances are obtained from $10^4$ realisations of an idealised survey with $N_{\text{eff}}$ independent $\omega$ bins.}
\label{fig:MultiStatComp}
\end{figure} 
Figure \ref{fig:MultiStatComp} compares the adjusted significances extracted for $M$ peaks with given raw significance for various small values of $M$. The agreement is very good indicating that using the integrated statistic to extract significances for multifrequency models is a valid way of analysing the data.

\onecolumngrid
\bibliography{correlation.bib}

\end{document}